\documentclass{aa}
\usepackage{graphicx,txfonts,natbib,color}
\bibpunct{(}{)}{;}{a}{}{,}

\begin{document}
\title{Eccentricity of radiative discs in close binary-star systems}
\author{F. Marzari\inst{1}, C. Baruteau \inst{2,3}, H. Scholl
  \inst{4} and P. Thebault \inst{5} }
\institute{
  Dipartimento di Fisica, University of Padova, Via Marzolo 8,
  35131 Padova, Italy
  \and
  DAMTP, University of Cambridge, Wilberforce Road, Cambridge CB30WA, United Kingdom
  \and Department of Astronomy and
  Astrophysics, University of California, Santa Cruz, CA 95064, USA
  \and Laboratoire Cassiop{\'e}e, Universit{\'e} de Nice Sophia
  Antipolis, CNRS,
  Observatoire de la C{\^o}te d'Azur, B.P. 4229, F-06304 Nice Cedex, France
  \and Observatoire de Paris, Section de
  Meudon,
  F-92195 Meudon Principal Cedex, France
}
\titlerunning{Radiative discs in close binaries}
\authorrunning{F. Marzari et al.}
\abstract 
%%  Context
{Discs in binaries have a complex behavior because of the
  perturbations of the companion star. 
Planetesimals growth and planet formation in binary-star 
systems both depend on the companion star parameters and on the 
properties of the circumstellar disc. An eccentric disc may significantly 
increase the impact velocity of planetesimals and therefore jeopardize 
the accumulation process.
}
%%  Aims
{We model the evolution of discs in close binaries including the effects of
  self-gravity and adopting different  prescriptions to model the 
  disc's radiative properties. We focus on
  the dynamical properties and evolutionary tracks of the discs.}
%% Methods
{We use the hydrodynamical code FARGO and we include in the energy
  equation heating  and cooling effects.  }
%% Results
{ Radiative discs have a lower disc eccentricity compared to 
  locally isothermal discs with same temperature profile.
  Its averaged value is about 0.05, and it is
  almost independent of the eccentricity of the binary orbit, in
  contrast to locally isothermal disc models. As a consequence, 
  we do not observe the formation of an internal elliptical low density 
  region as in locally isothermal disc models. However, the disc
  eccentricity depends on the disc mass through the opacities. 
Akin to locally isothermal disc models, self-gravity forces the 
disc's longitude of pericenter to librate about a fixed orientation 
with respect to the binary apsidal line ($\pi$).
}
%% Conclusions
{ The disc's radiative properties play an important role in the
  evolution of discs in binaries. 
  A radiative disc has an overall shape and internal structure that 
  are significantly different compared to a locally isothermal disc with same temperature profile.  
   This is an important 
  finding both for describing the evolutionary track of the disc 
  during its progressive mass loss,  
  and for planet formation since the internal structure of the
  disc is relevant for planetesimals  growth in binary systems. The non-symmetrical distribution of mass in these discs
  causes large eccentricities for planetesimals that may affect {their growth}.}
\maketitle

% ====================== %
\section{Introduction} 
% ====================== %
\label{intro}
Protoplanetary discs are known to exist in pre-main-sequence binary
star systems through direct imaging \citep{koe, sta, rod} and spectral
energy distributions \citep{ghez, pra, mon}. \cite{goz} have recently
claimed that as much as 80\% of all binary star systems in the $\sim$
1 Myr old Orion Nebula Cluster might have an active accretion disc.
\cite{tri} showed that the incidence of debris discs in binary systems
is not that different than that for single stars. These observations
seem to suggest that the ground for planet formation in binary star
systems is present even if, at subsequent stages, the gravitational
perturbations by the companion star may negatively impact the growth
process. \cite{bona,mugra} have shown that the frequency of planets in
binaries is not very different from that around single stars. However,
this frequency critically depends on the binary separation and close
binaries appear to be less favorable environments for planet
formation. Only a few planets are presently known in tight systems
(GI86, HD41004 and $\gamma$ Cephei) and it is expected that when the
binary separation is in the range 20-100 AU the gravitational
influence of the companion star must have influenced the planet
formation process, affecting either the disc evolution or the
planetesimal accumulation process \citep{desibar}.

Of particular importance are the initial stages of planetesimal
evolution when the mutual impact velocities must remain low to allow
mass growth despite the perturbations by the companion star. Previous
studies
\citep[e.g.,][]{mascho,theb06,theb08,theb09,paard08,xie09,paard10}
have shown that the combination of gas drag force and secular
perturbations by the secondary star has strong effects on planetesimal
orbits, leading to pericenter alignment and to an equilibrium
distribution for the eccentricity of small planetesimals. Since the
magnitude of the gas drag force depends on the planetesimals size, the
variation of eccentricity and pericenter longitude with particle size
may well inhibit the formation of larger bodies by exciting large
mutual impact velocities.

One question that needs to be addressed is to what extent the disc's
perturbation due to the companion star
\citep[e.g.,][]{lub91a,lub91b,kle08} affects the distributions of
eccentricity and longitude of pericenter for planetesimals of
different size.  Models in which the planetesimals orbital evolution
was computed along with the time evolution of the disc
\citep{kn08,paard08} have shown that the disc eccentricity may
strongly influence the dynamics of small planetesimals by introducing
radial gas drag forces and non-axisymmetric components in the gravity
field of the disc. This additional perturbing force may act either in
favor or against planetesimal accretion by affecting the relative
impact velocities between the bodies. An in depth analysis is needed
to explore the strength of these perturbations. This first requires
investigating the impact of the companion star on the disc's
eccentricity.  We focus on relatively close binary star systems with
large mass ratios and potentially large eccentricities which,
according to \cite{duma}, populate the peak of the frequency
distribution in our neighborhood.  Modelling the response of a
protoplanetary disc to the gravitational perturbations by a secondary
(companion) star requires by necessity a numerical approach, since it
is difficult to predict analytically the disc shape and evolution with
time.

In a previous paper (\cite{mabash}, hereafter Paper I) we used the
numerical code FARGO to model the time evolution of a two-dimensional
(2D) circumstellar disc in close binary star systems, including the
effects of disc self-gravity. We focused on massive discs, with an
initial mass equal to $0.04 M_{\odot}$, as well as a large binary's
mass ratio ($\mu = M_s/M_p = 0.4$). This value is statistically the
most frequent among the binary systems observed so far \citep{duma}.
We found that self-gravity significantly affects the increase in the
disc eccentricity induced by the companion perturbations, and also
influences the orientation of the disc relative to the binary
reference frame by causing libration instead of circulation.  We also
sampled different values of the binary's eccentricity, $e_b$, ranging
from 0 to 0.6. The main findings of paper I were the following:
\begin{itemize}
\item self-gravity plays a significant role in shaping the disc.  The
  dynamical eccentricity $e_d$ (defined in Sect. 2.3) of the disc
  typically ranges from $\sim 0.05$ to $\sim 0.15$, depending on the
  binary eccentricity, $e_b$. It is smaller than in models without
  self gravity,
  \item $e_d$ is inversely proportional to $e_b$, with the case of 
  a circular binary ($e_b = 0$) being the most perturbing configuration,
\item the disc orientation $\omega_d$ (defined in Sect. 2.3)
librates around $\pi$, while it was circulating in the absence of self gravity,
\item an eccentric low-density region develops in the inner disc parts
  because of the large eccentricity and aligned pericenters of the gas
  streamlines there.
\end{itemize}

The results of paper I have been obtained assuming a locally
isothermal equation of state for the gas, wherein the initial radial
profile of the temperature remains constant in time, its value being
set by the choice for the disc aspect ratio $h = H/r$, with $H$ the
pressure scale height.  This approximation is well suited in disc
regions where radiation cooling is efficient and the gas is optically
thin. However, in particular in the initial stages of their evolution,
discs may be optically thick and a more detailed treatment of the
energy balance is required. In addition, when it passes at its
pericenter, the secondary star triggers spiral shocks that may
generate local strong shock and compressional heating, which may
violate the local isothermal approximation. The propagation of these
shock waves may also be significantly altered in radiative discs. In
this paper, we focus on how the disc eccentricity and orientation
depend on the disc radiative properties. Our approach is the
following. We first examine how our previous results on locally
isothermal discs depend on the choice for the (fixed) temperature
profile of the disc. We then consider disc models with 
a radiative energy equation, for which we find that 
the averaged disc's eccentricity in its inner parts is smaller compared to 
locally isothermal disc models with similar temperature profile.
 We finally discuss the impact of our results in terms of planetesimals 
 dynamics.

% ===========================
\section{Model description}
% ===========================
\label{sec:model}

% ---------------------------------------
\subsection{Code}
% ---------------------------------------
Two-dimensional hydrodynamical simulations were carried out with
the ADSG version\footnote{See: {\texttt http://fargo.in2p3.fr}} of the
code FARGO. The code solves the hydrodynamical equations on a polar
grid, and it uses an upwind transport scheme along with a harmonic,
second-order slope limiter \citep{vl77}. The ADSG version of the FARGO
code includes an adiabatic energy equation and a self-gravity module
based on fast Fourier transforms \citep{bm08b}. Heating and cooling
source terms have been implemented in the energy equation as described
in Sect.~\ref{sec:setup}. The specificity of the FARGO algorithm is to
use a change of rotating frame on each ring of the grid, which
increases the timestep significantly.

Results of simulations are expressed in the following code units: the
mass unit is the mass of the primary star, $M_p$, which is taken
equal to $1 M_{\odot}$. The length unit is set to 1 AU, and the orbital
period at 1 AU is $2\pi$ times the code's time unit.

% ---------------------------------------
\subsection{Physical model and numerical setup}
% ---------------------------------------
\label{sec:setup}
We adopt a 2D disc model in which self-gravity and an energy equation
are included (unless otherwise stated). The hydrodynamical equations
are solved in a cylindrical coordinate system $\{r,\varphi\}$ centered
onto the primary star, with $r \in [0.5 {\,\rm AU} - 15 {\,\rm AU}]$ and
$\varphi \in [0,2\pi]$. The grid used in our calculations has $N_{\rm
  r} = 256$ radial zones and $N_{\rm s} = 512$ azimuthal zones, and a
logarithmic spacing is used along the radial direction. The frame
rotates with the Keplerian angular velocity at the binary's semi-major
axis, and the indirect terms accounting for the acceleration of the
primary due to the gravity of the secondary and of the disc are
included. 
\\
%%%%%%%%%%  
\par\noindent\emph{Binary parameters---}
%%%%%%%%%%
Throughout this study, we adopt as standard model a binary system
where the secondary star has a mass $M_{\rm s} = 0.4M_{\odot}$. The 
binary is held on a fixed eccentric orbit with semi-major axis $a_{\rm b} = 30$ AU
and eccentricity $e_{\rm b} = 0.4$, corresponding to an orbital period
$\approx 134$ yr.
\\
%%%%%%%%%%  
\par\noindent\emph{Energy equation---}
%%%%%%%%%%
Since the purpose of this work is to examine the impact of an energy
equation on the disc's response to the periodic passages of a close
companion, we carried simulations including either an energy equation,
or a locally isothermal equation such that the initial
temperature profile remains constant in time. In all cases, the disc
verifies the ideal gas law,
\begin{equation}
p = {\cal R} \Sigma T / \mu,
\end{equation}
where $p$ and $T$ denote the vertically-integrated pressure and
temperature, respectively, $\Sigma$ is the mass surface density,
${\cal R}$ is the ideal gas constant, and $\mu$ is the mean molecular
weight, taken equal to 2.35. When included, the energy equation takes
the form:
 \begin{equation}
    \frac{\partial e}{\partial t} + {\bf \nabla} \cdot (e {\bf v}) 
    = -p {\bf \nabla} \cdot {\bf v} 
    + Q^{+}_{\rm visc}
    - Q^{-}_{\rm cool}
    + \lambda e \nabla^2 \log(p/\Sigma^{\gamma})
  \label{eqnenergy}
  \end{equation}
  where $e$ = $p / (\gamma-1)$ is the thermal energy density, $\gamma$
  is the adiabatic index, and ${\bf v}$ denotes the gas velocity.  We
  take $\gamma=1.4$ throughout this study.  In Eq.~(\ref{eqnenergy}),
  $Q^{+}_{\rm visc}$ denotes the viscous heating. We use both a
  constant shear kinematic viscosity, $\nu = 10^{-5}$ in code units,
  and a von Neumann-Richtmyer artificial bulk viscosity, as described
  in \cite{zeus}, where the coefficient $C_2$ is taken equal to 1.4
  ($C_2$ measures the number of zones over which the artificial
  viscosity spreads out shocks). The cooling source term in
  Eq.~(\ref{eqnenergy}), $Q^{-}_{\rm cool}$, is taken equal to $2
  \sigma_{\rm SB} T_{\rm eff}^4$, where $\sigma_{\rm SB}$ is the
  Stefan-Boltzmann constant, and $T_{\rm eff}$ is the effective
  temperature \citep{Hubeny90},
  \begin{equation}
  T^4_{\rm eff} = T^4 / \tau_{\rm eff},
  \end{equation}  
  with effective optical depth
  \begin{equation}
  \tau_{\rm eff} = \frac{3\tau}{8} + \frac{\sqrt{3}}{4} + \frac{1}{4\tau}.
  \end{equation} 
  The vertical optical depth, $\tau$, is approximated as $\tau =
  \kappa\Sigma / 2$, where for the Rosseland mean opacity, $\kappa$,
  the formulae in \cite{beli} are adopted. Following \cite{pbk11}, we
  also model thermal diffusion as diffusion of the gas entropy, $s$,
  defined as $s = {\cal R}(\gamma-1)^{-1} \log(p /
  \Sigma^{\gamma})$. This corresponds to the last term in the
  right-hand side of Eq.~(\ref{eqnenergy}), where $\lambda$ is a
  constant thermal diffusion coefficient. Throughout this study, we
  adopt $\lambda=10^{-6}$ in code units.
  \\
%%%%%%%%%%  
\par\noindent\emph{Initial conditions---}
%%%%%%%%%%
The disc is initially axisymmetric and 
the angular frequency $\Omega(r)$ about the primary star is
computed taking into account the radial acceleration due to the pressure gradient,
the gravitational acceleration due to the primary star only and 
the self--gravity of the disk. 
The
initial gas surface density, $\Sigma_0$, is set as $\Sigma_0(r) =
\Sigma_0(1\,{\rm AU}) \times (r/1\,{\rm AU}) ^{-1/2}$ with
$\Sigma_0(1\,{\rm AU}) = 2.5 \times 10^{-4}$ in code units. Assuming
the mass of the primary star is $1 M_{\odot}$, this corresponds to
setting the disc surface density at 1 AU to $\approx 2.2\times 10^{3}$ g
cm$^{-2}$. Beyond 11 AU, $\Sigma_0(r)$ is smoothly reduced to a floor
value, $\Sigma_{\rm floor} = 10^{-9} = 4\times
10^{-6}\,\Sigma_0(1\,{\rm AU})$, using a Gaussian function with
standard deviation equal to $0.4$ AU. To prevent numerical
instabilities caused by low density values or by steep density
gradients near the grid's outer edge, the gas density in each grid
cell is reset to $\Sigma_{\rm floor}$ whenever it becomes smaller than
this floor value \citep[][Paper I]{kle08}. Similarly, we adopt a floor
value for the thermal energy density, $e_{\rm floor} = 10^{-18}$.  Our
choice for this parameter is conservative, and we checked that with 
a larger $e_{\rm floor}$, the modelled disc behaved similarly.  The
initial gas temperature, $T_0$, is taken proportional to $r^{-1}$:
$T_0 = T_0(1\,{\rm AU}) \times (r/1\,{\rm AU}) ^{-1}$, where the value
for $T_0(1\,{\rm AU})$, which is taken as a free parameter in our
study, will be specified below. Writing the pressure scale height $H$
as $H = c_{\rm S} \Omega^{-1}_{\rm K}$, where $c_{\rm S}$ denotes the
sound speed and $\Omega_{\rm K}$ the Keplerian angular frequency, the
disc temperature can be conveniently related to the disc's aspect
ratio, $h = H/r$. In our code units, where ${\cal R}/\mu = 1$, this
yields $T_0(1\,{\rm AU}) = 630\,{\rm K} \times \{ h(1\,{\rm AU}) /
0.05 \}^{2}$.
\\
%%%%%%%%%%  
\par\noindent\emph{Boundary conditions---}
%%%%%%%%%%
By default, an outflow zero-gradient boundary condition is adopted  
at both the grid's inner and outer edges. The azimuthal velocity is set 
to its initial, axisymmetric value. No mass therefore flows back into the system, 
and the disk mass declines with time. The impact of the inner boundary condition 
on our results is examined in Sect.~\ref{sec:boundary}.

% ---------------------------------------
\subsection{Notations}
% ---------------------------------------
We will make use of the following quantities. We denote by $e_{\rm d}$
and $\varpi_{\rm d}$ the disc's eccentricity and perihelion longitude,
respectively. They are defined as in \citet{kle08, pine, mabash}:
\begin{equation}
  e_{\rm d} = M^{-1}_{\rm d} \times \iint e(r,\varphi) \Sigma(r,\varphi) r dr d\varphi
\end{equation}
and
\begin{equation}
  \varpi_{\rm d}  =  M^{-1}_{\rm d} \times \iint \varpi(r,\varphi) \Sigma(r,\varphi) r dr d\varphi
\end{equation}
where $M_{\rm d}$ denotes the disc's mass, and $e$ and $\varpi$ are
the eccentricity and pericenter longitude of each grid cell,
respectively, assuming that the local position and velocity vectors
uniquely define a 2-body Keplerian orbit.

% ====================== %
\section{Locally isothermal disc models}
% ====================== %
\label{sec:isoth}
Before examining the impact of an energy equation on the disc's
response to the tidal perturbations by the secondary star, we adopt in
this section the simpler case for a locally isothermal equation of
state, where the initial (axisymmetric) profile of the disc
temperature remains fixed. This first step will help us analyze the
more complex situation of a disc whose temperature evolves in time due
to radiative cooling and various sources of heating, including that
arising from the shock waves induced by the secondary.

% ---------------------------------------
\subsection{Disc's eccentricity and surface density profiles}
\begin{figure*}[hpt]
  \centering\resizebox{\hsize}{!}  {
    \includegraphics{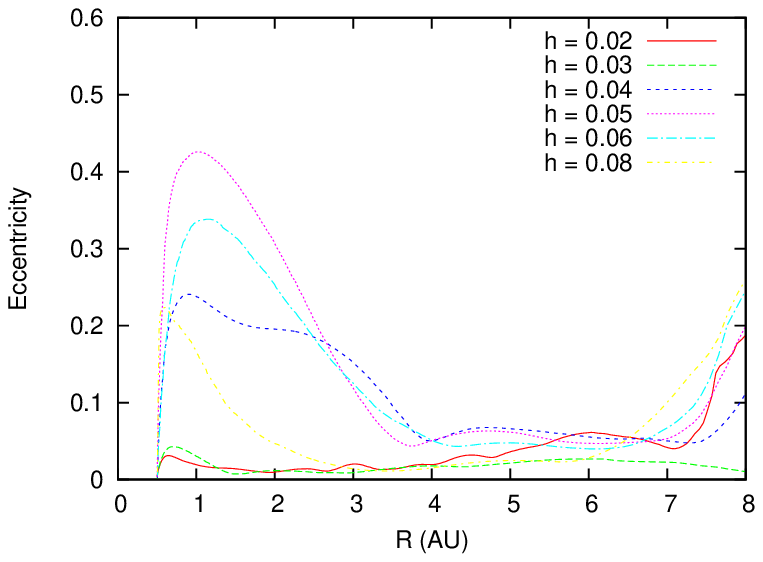}
    \includegraphics{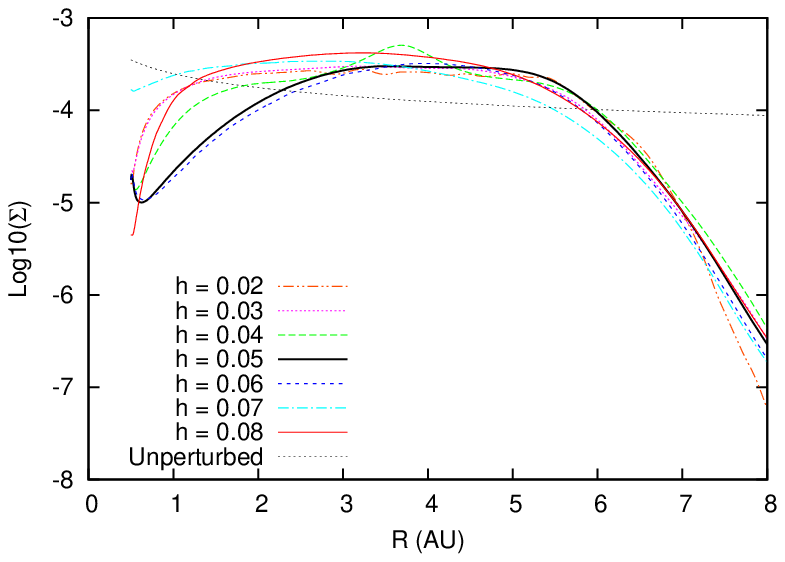}
   } 
   \centering\resizebox{\hsize}{!}  {
    \includegraphics{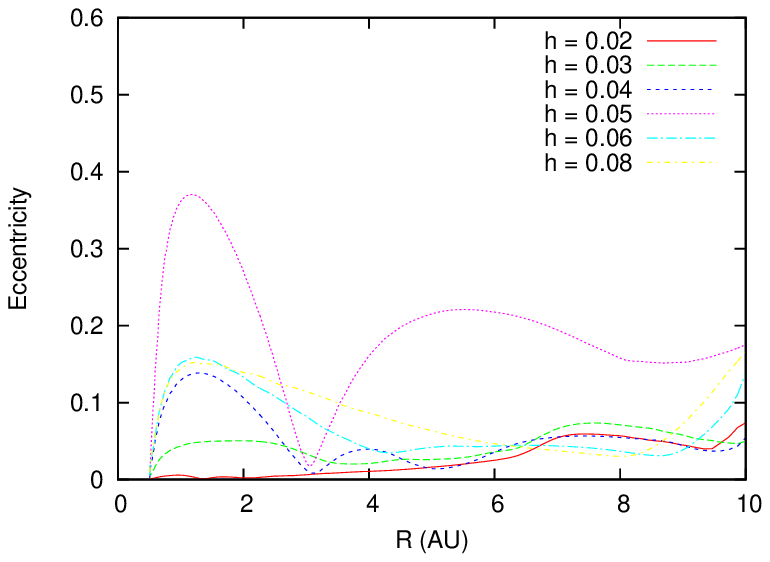}
    \includegraphics{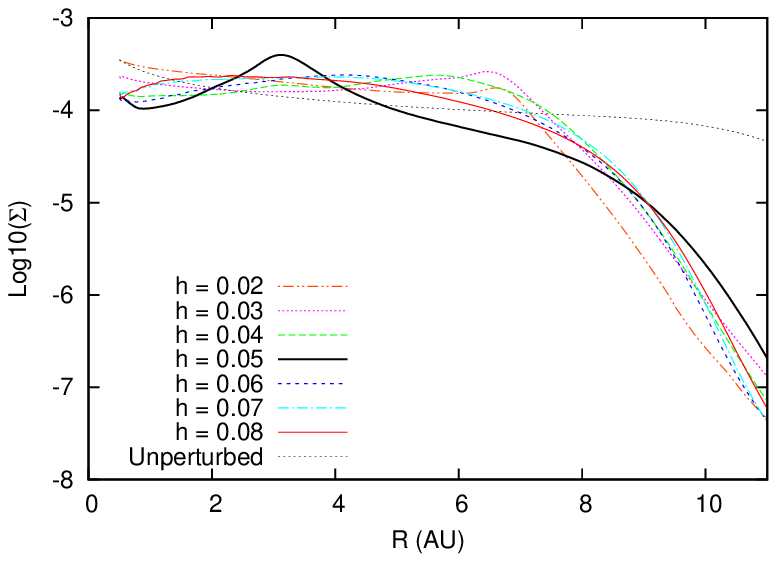}
   }
   \caption{\label{f_iso}Azimuthally- and time-averaged profiles of
     the disc's eccentricity (left column) and surface density
     ($\Sigma$, right column) obtained with the series of locally
     isothermal disc models described in
     Sect.~\ref{sec:isoth}. In the density profile plots the 
     initial, unperturbed profile is shown as reference.  Results are shown at 60 orbits of the
     secondary, and time-averaging is done over 5 orbits. Several 
     values of the disc's aspect ratio at 1 AU are considered 
     and two secondary-to-primary mass ratios: $q=0.4$ (upper
     panels) and $q=0.1$ (lower panels).  }
\end{figure*}

% ---------------------------------------
\label{sec:31}
We carried out a series of simulations using a range of values for the
disc's aspect ratio at 1 AU, $h(1\,{\rm AU})$. This comes to varying
the disc's temperature at the same location. Other disc and binary
parameters are as described in Sect.~\ref{sec:setup}. We consider two
different values for the secondary-to-primary mass ratio: $q=0.4$ (our
fiducial value), and $q = 0.1$. Results of simulations with $q = 0.4$
are shown in the upper panels in Fig.~\ref{f_iso}, and those with $q=0.1$
in the lower  panels. Azimuthally- and time-averaged profiles of the
disc's eccentricity and surface density are displayed in the left and
right columns of Fig.~\ref{f_iso}, respectively. Profiles are displayed
after $\sim 60$ orbits of the secondary, and time averaging is done
over 5 orbits. By checking the time evolution of the disc's eccentricity
profile, we find that a steady state is reached
after typically 20 binary revolutions with $q=0.4$ for all values of
$h$, and after 40 revolutions for $q=0.1$. The value of $h(1\,{\rm AU})$ is
indicated in each panel (and simply denoted by $h$).

We are primarily interested in the disc's averaged eccentricity in its
inner parts, below the truncation radius located at $r \sim 5-6$ AU 
for $q=0.4$, where planet formation is more likely to occur. As shown in
\cite{mabash}, the truncation radius of the disc closely matches the
critical limit for orbital stability due to the secular perturbations
of the companion star \citep{holwi}. From Fig.~\ref{f_iso}, it is clear
that the disc eccentricity increases from $r \sim 4$ AU downwards,
peaks at $r \lesssim 1$ AU, then decreases towards the location of the
grid's inner edge, where the boundary condition imposes zero
eccentricity. Interestingly, we see that the averaged peak
eccentricity of the disc (reached at $r \lesssim 1$ AU) increases with
$h$ up to $h = 0.05$ and decreases beyond this value. The same
behavior is obtained with both values of $q$.  This behavior may be 
interpreted as follows. The disc's density perturbation due to the 
secondary decreases with increasing disc aspect ratio ($h$ or,
equivalently, increasing sound speed). In the limit of large aspect ratios, 
the disc's eccentricity thus decreases with increasing $h$. As $h$ 
decreases, the disc's perturbed density increases, but shock waves 
become more tightly wound, and have to travel a longer distance 
before reaching the disc's inner parts. Being then more prone to viscous 
damping, shock waves become less and less efficient at depositing angular 
momentum in the disc's inner parts, where the eccentricity remains small. 
This explains why when decreasing $h$, the averaged eccentricity of the 
disc's inner parts first reaches a maximum and then decreases.

We can see in Fig.~\ref{f_iso} that the radial dependence of the
density and eccentricity profiles reasonably match each other, as
expected. The general trend is that the larger the disc's eccentricity
profile, the smaller the surface density profile. Note that this is
more visible for $q=0.4$, where the density's perturbation by the
secondary takes larger values than for $q=0.1$. The significant
decrease in the disc's density profile in the disc inner parts
(typically below $\sim 2$ AU) takes the form of an elliptic inner
hole, which has been observed in a number of previous numerical
studies \citep[e.g.,][Paper I]{kle08}.

Our results concerning the dependence of the disc's eccentricity and
density profiles with $h$ differ from those of \cite{kle08}, but this
depends on intrinsic differences in the models.  In particular, we
consider more massive discs and we include the effects of
self-gravity which proved to be efficient in affecting the disc
evolution.

%FFFFFFFFFFFFFFFFF
%FFFFFFFFFFFFFFFFF

% ---------------------------------------
\subsection{Fourier analysis of the density distribution}
% ---------------------------------------
To provide some insight into the dependence of the disc's eccentricity
and density profiles with varying temperature, we examine in this
paragraph the Fourier components of the disc's surface
density. Fig.~\ref{f_fou} displays the instantaneous amplitude of the
Fourier mode coefficients with azimuthal wavenumber $1 \leq m \leq
5$. Results are shown at 2550 yr, that is after $\approx 20$ orbits of
the secondary, when the disc is truncated at about 7-8 AU. We compare
the profiles obtained for previous series of locally isothermal disc
models for $q=0.4$, with $h = 0.02$ (left panel) and $h = 0.06$ (right
panel).  The secondary star is half-way between the pericenter and
apocenter at this particular point in time.

From Fig.~\ref{f_fou}, it is clear that the amplitude of Fourier mode
coefficients decreases with increasing $m$, and that the $m=1$ 
mode prevails. The run with $h = 0.06$ displays larger
mode amplitudes throughout the disc's inner parts, which accounts for
the larger disc eccentricity obtained in this case. Also, the large
amplitude of the $m=1$ mode below $r \approx 1$ for $h=0.06$ is
directly associated to the presence of an elliptic inner hole, whose
presence has already been pointed out in the upper panels in
Fig.~\ref{f_iso}.

%FFFFFFFFFFFFFFFFF
\begin{figure}[hpt]
  \includegraphics[width=\hsize]{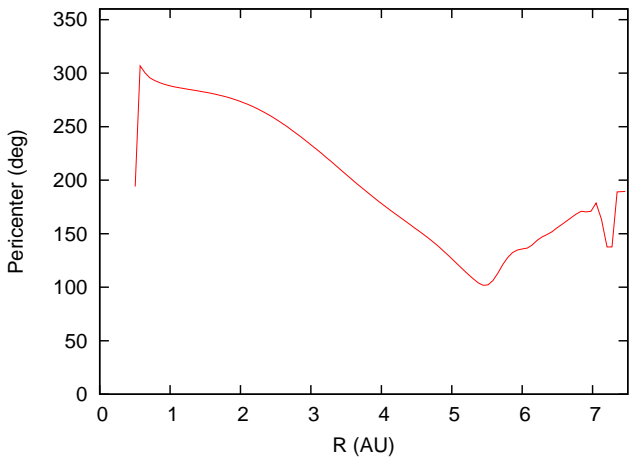}
  \caption{\label{f_ome}Azimuthally-averaged profile of the disc's
    pericenter in the locally isothermal disc model with $h = 0.05$
    and $q = 0.4$. It is computed after 100 orbits of the secondary
    star. }
\end{figure}
%FFFFFFFFFFFFFFFFF

%FFFFFFFFFFFFFFFFF
\begin{figure*}[hpt]
   \centering\resizebox{\hsize}{!}  {
     \includegraphics[angle=-90]{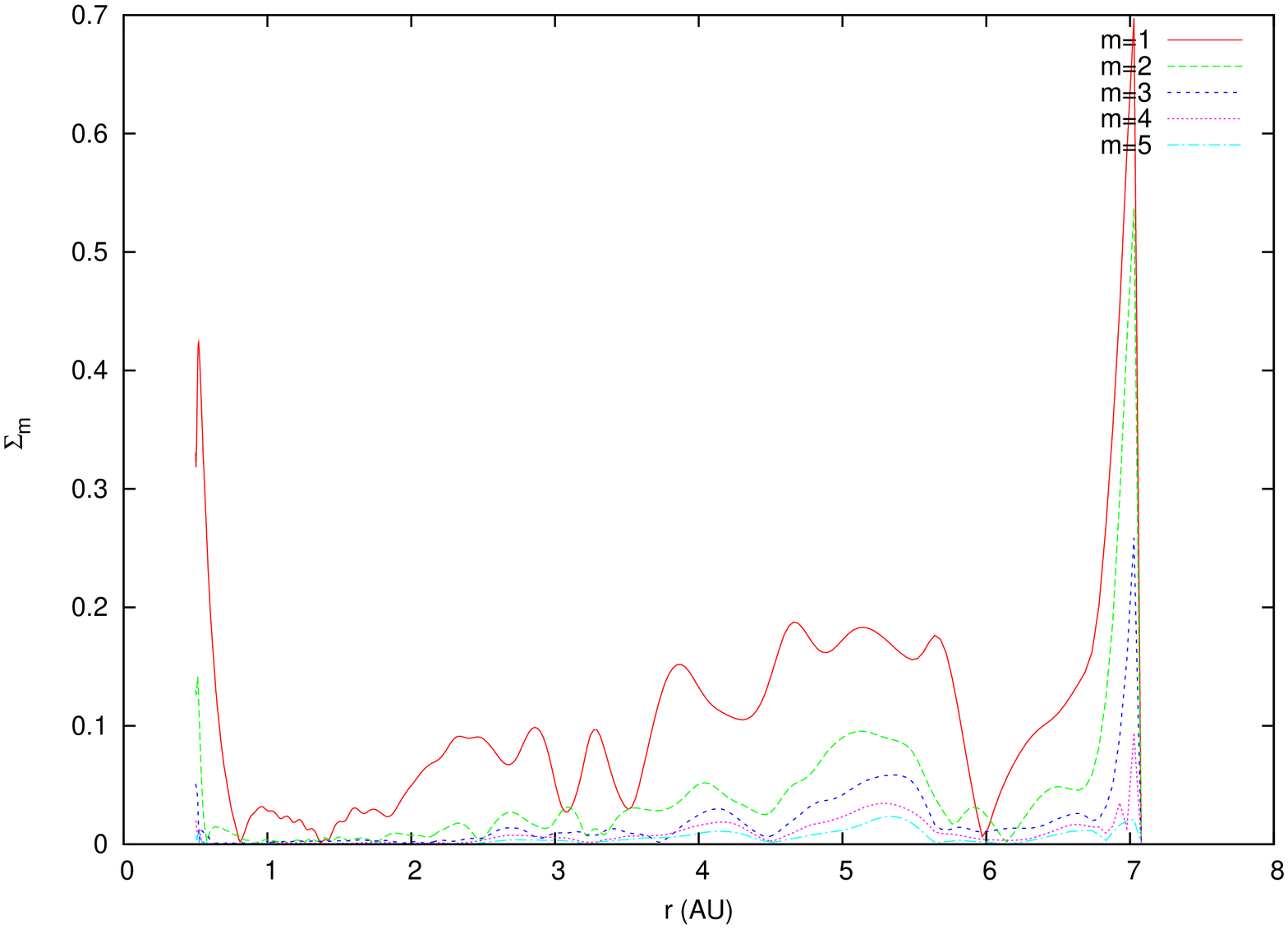}
     \includegraphics[angle=-90]{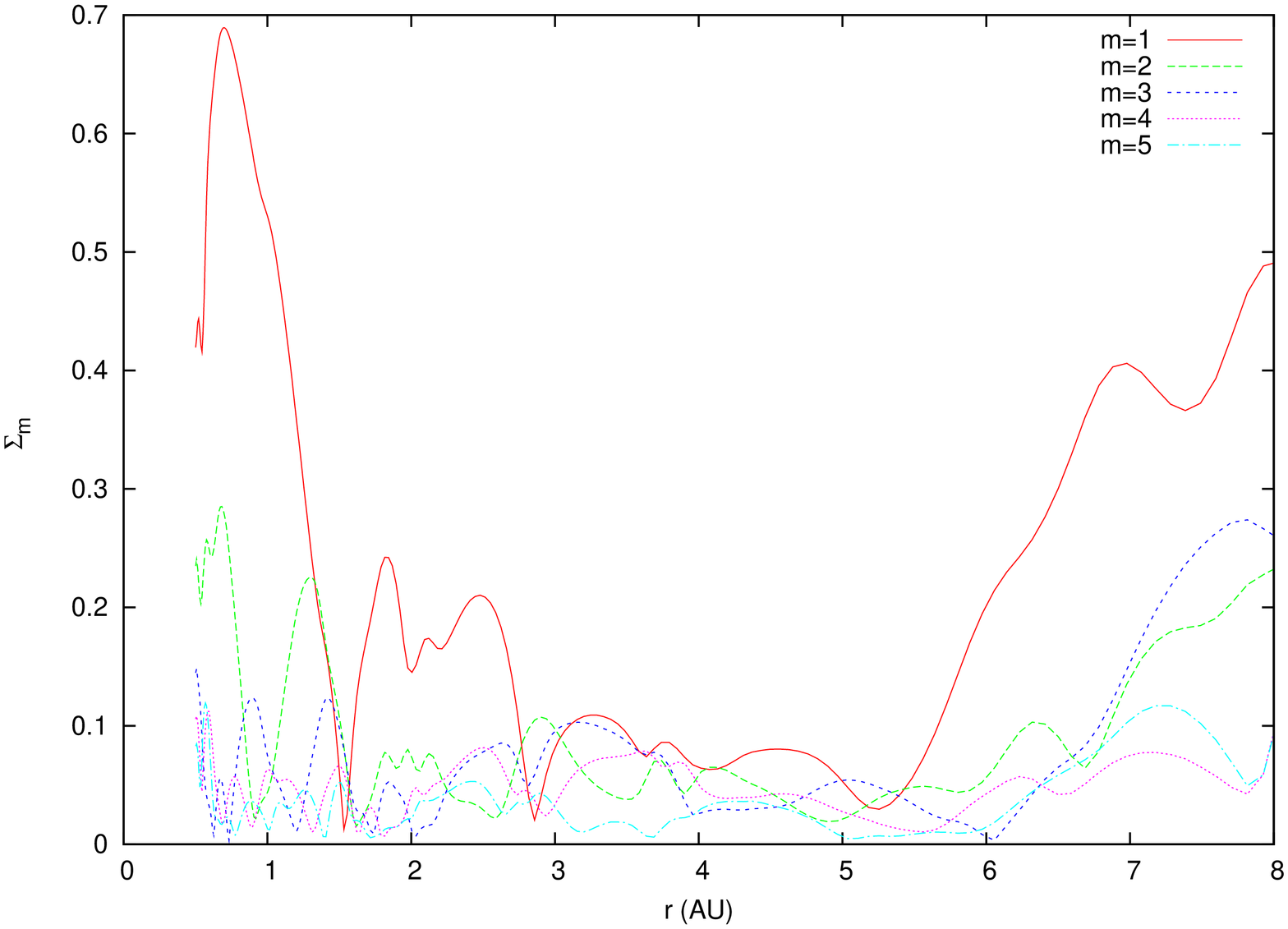}
   }
   \caption{\label{f_fou}Instantaneous Fourier components of the
     disc's surface density for the locally isothermal disc models of
     Sect.~\ref{sec:isoth} with $h = 0.02$ (left panel) and $h = 0.06$
     (right panel), respectively. Results are shown at $t=2550$ yr
     (that is, after about 20 orbits of the secondary), and are
     obtained with $q=0.4$. The azimuthal wavenumber for each Fourier
     component is indicated in the top-right corner of each panel.}
\end{figure*}
%FFFFFFFFFFFFFFFFF

% ====================== %
\section{Radiative discs models}
% ====================== %
In this section, we describe the results of our hydrodynamical
simulations that include an energy equation. Our aim is to assess the
impact of the energy equation on the disc's averaged eccentricity.

% ---------------------------------------
\subsection{Disc eccentricity}
% ---------------------------------------
\label{sec:rde}
%FFFFFFFFFFFFFFFFF
\begin{figure*}[hpt]
  \centering\resizebox{\hsize}{!}  {
    \includegraphics{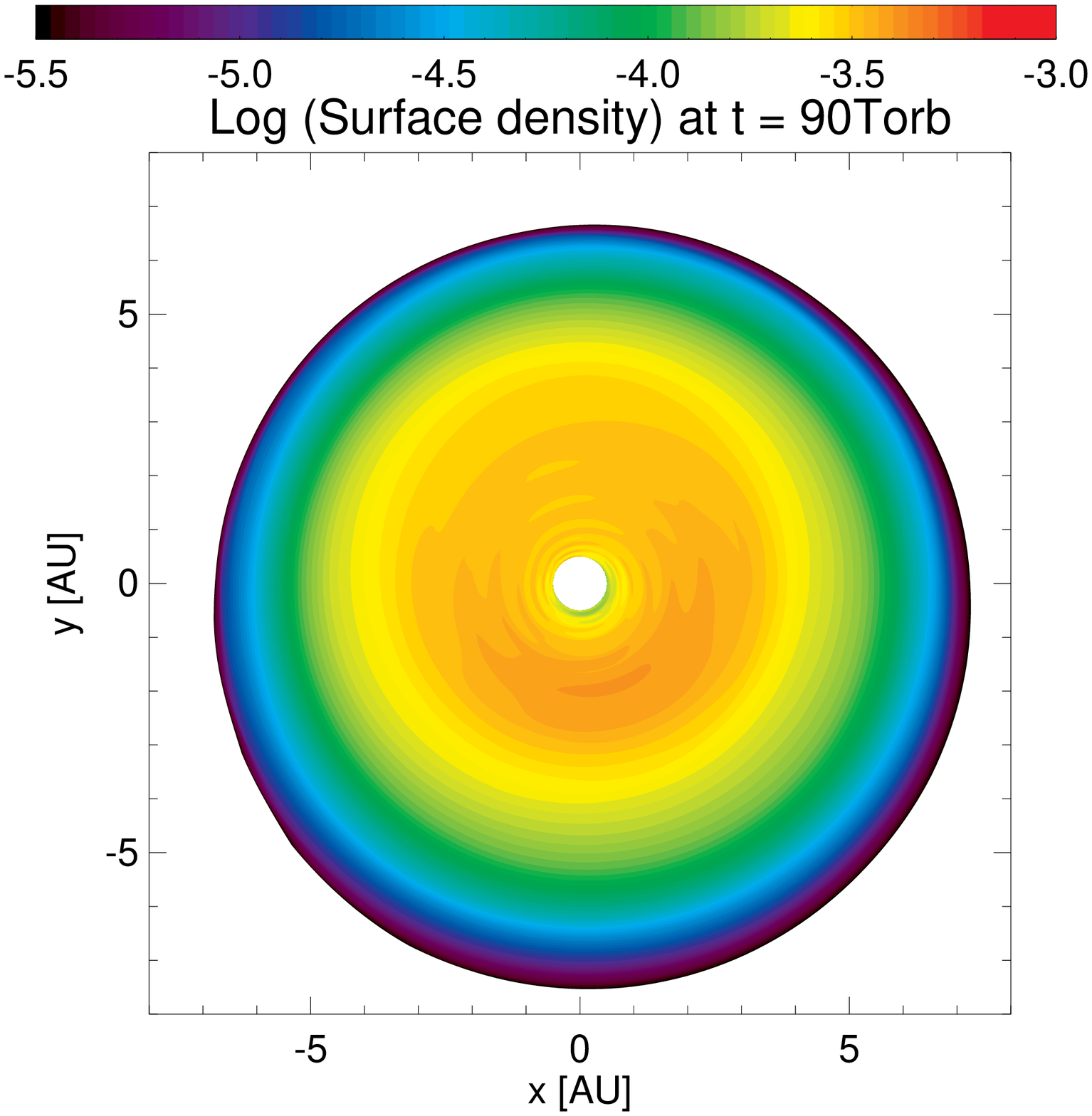}
    \includegraphics{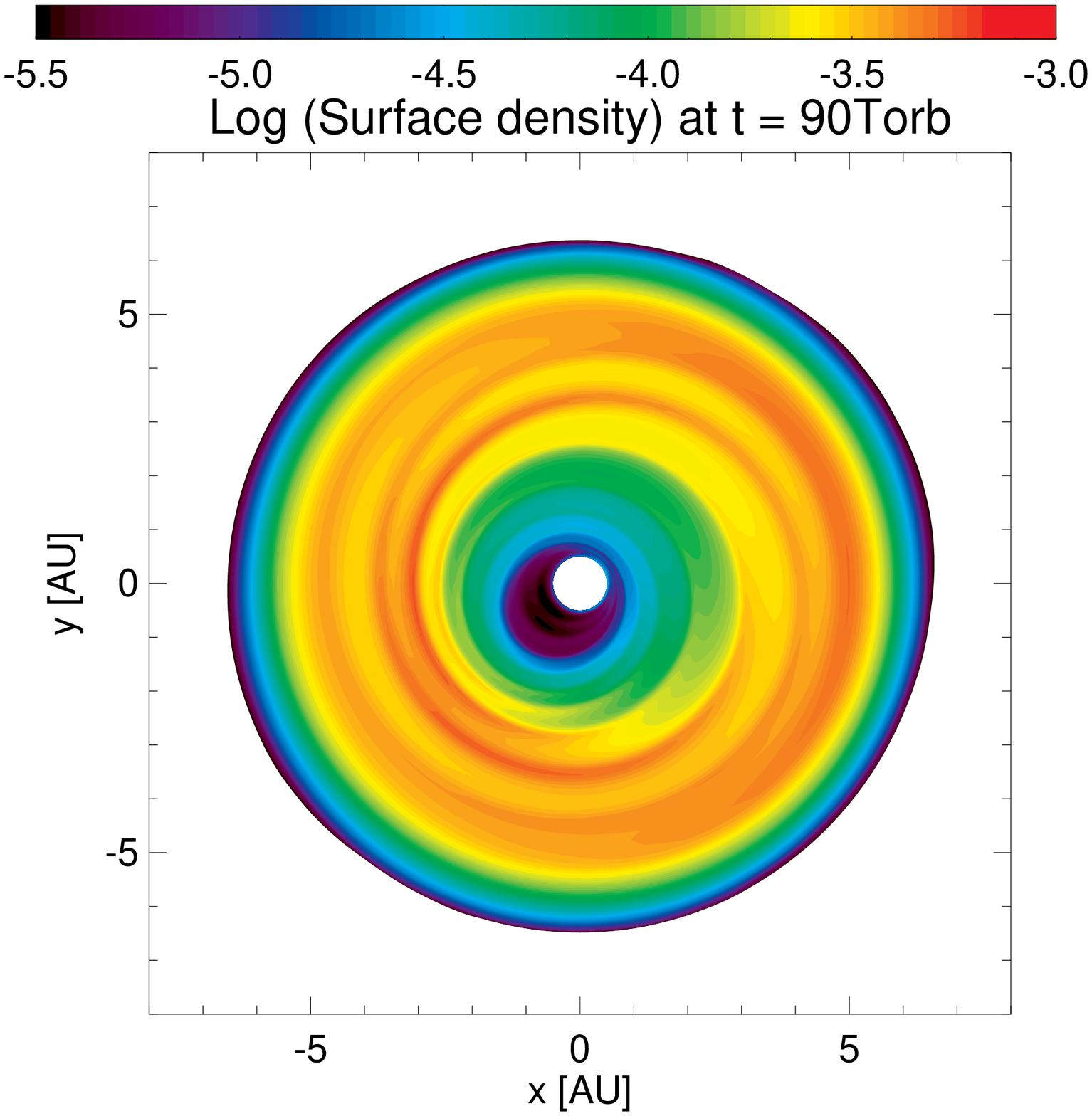}
   } 
   \centering\resizebox{\hsize}{!}  {
    \includegraphics{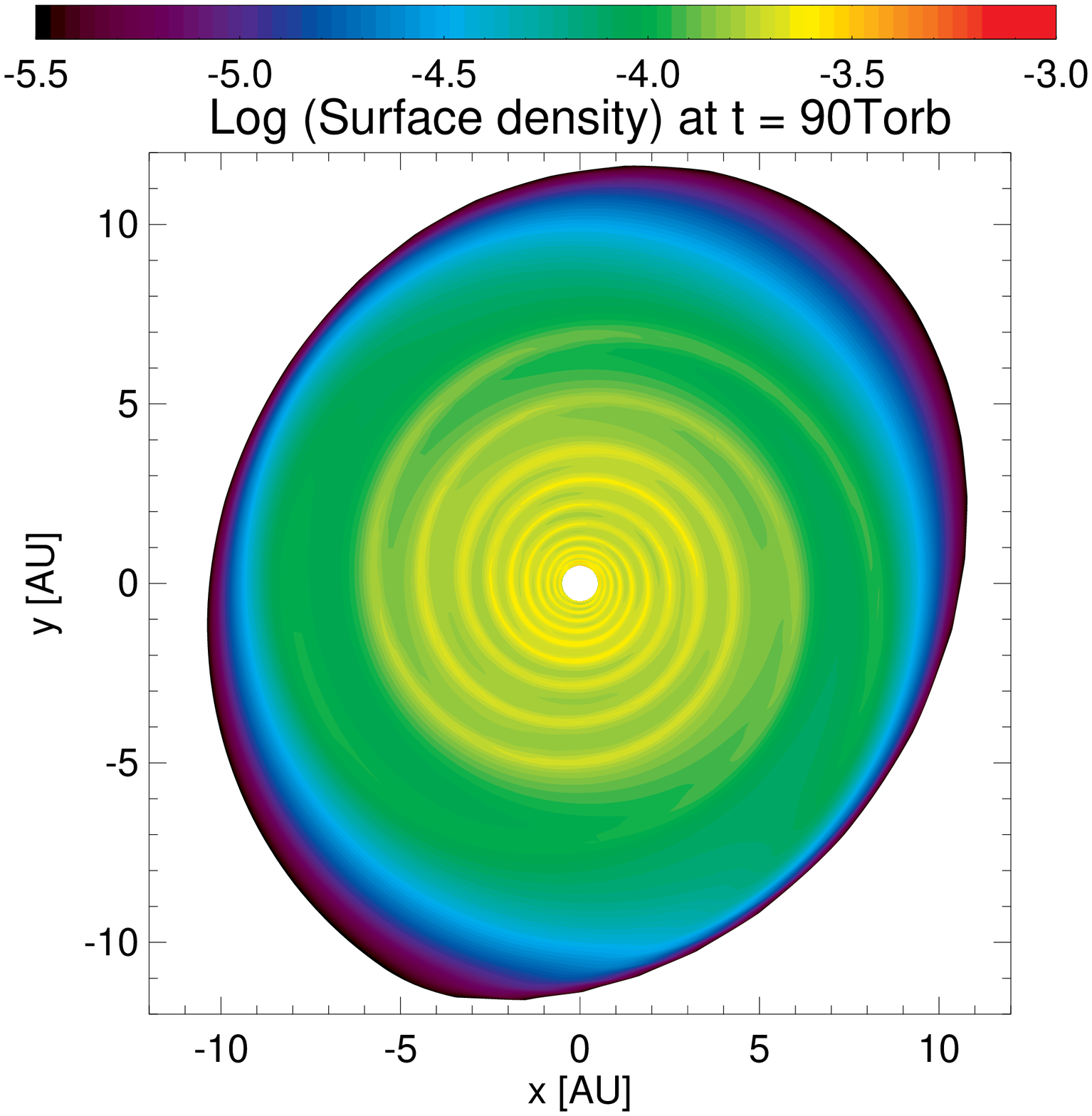}
    \includegraphics{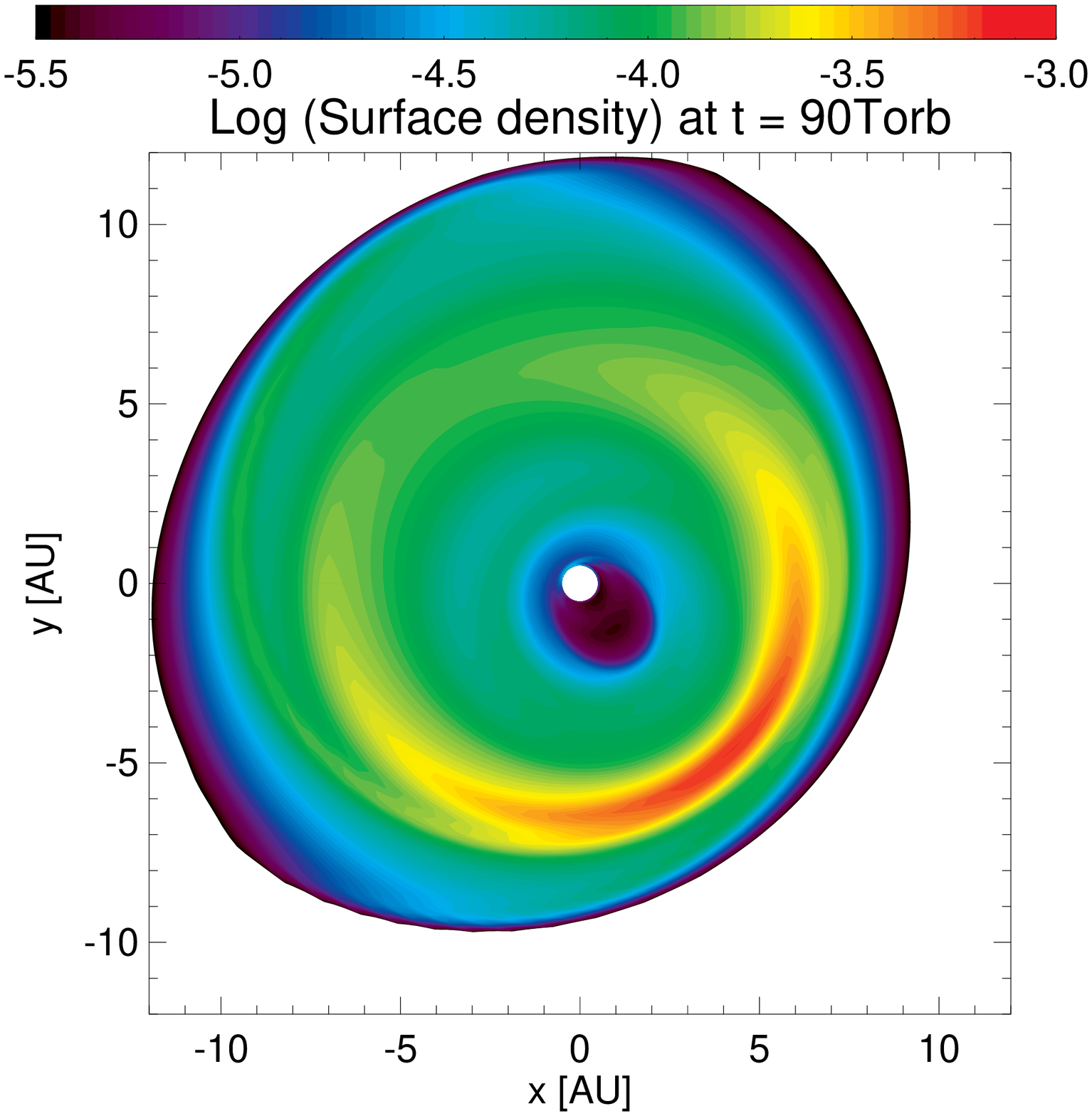}
   }
   \caption{\label{f_2D}Contours of the gas density after 90 binary
     revolutions. Results with an energy equation are shown in the
     left part of this figure, and those with a locally isothermal
     equation of state with  fixed temperature profile are shown
     in the right part. The upper plots are for $e_b=0.4$ and the
     lower plots for $e_b=0.0$. The same initial aspect ratio
     ($h=0.05$) is used in these simulations. }
\end{figure*}
%FFFFFFFFFFFFFFFFF

\begin{figure}[hpt]
  \includegraphics[width=\hsize]{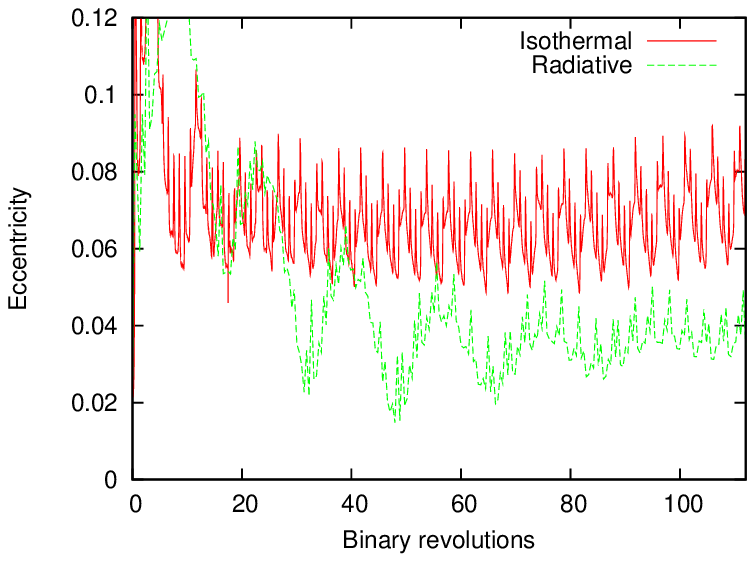}
  \includegraphics[width=\hsize]{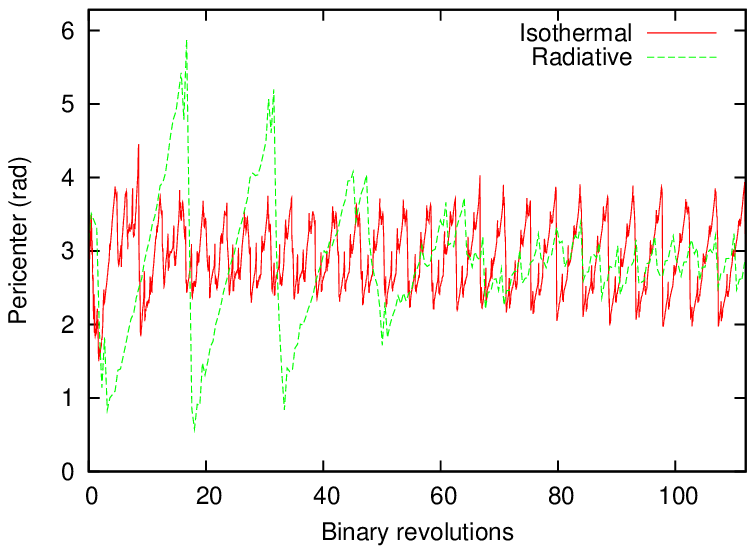}
  \caption{\label{f_eperi} Time evolution of the averaged disc 
  eccentricity $e_d$ (upper panel) and pericenter 
  longitude $\varpi_d$ (lower panel) for the locally isothermal 
  and radiative runs of Sect.~\ref{sec:rde}.}
\end{figure}

%FFFFFFFFFFFFFFFFF
\begin{figure*}[hpt]
  \centering\resizebox{\hsize}{!}  {
    \includegraphics{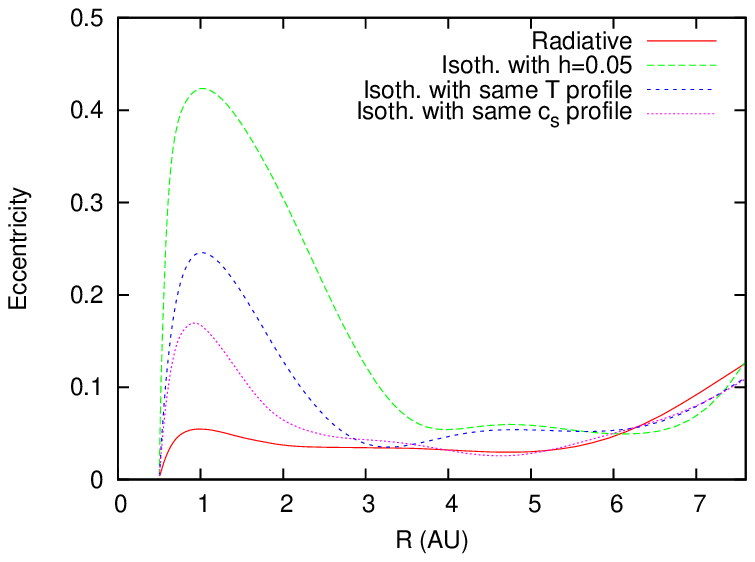}
    \includegraphics{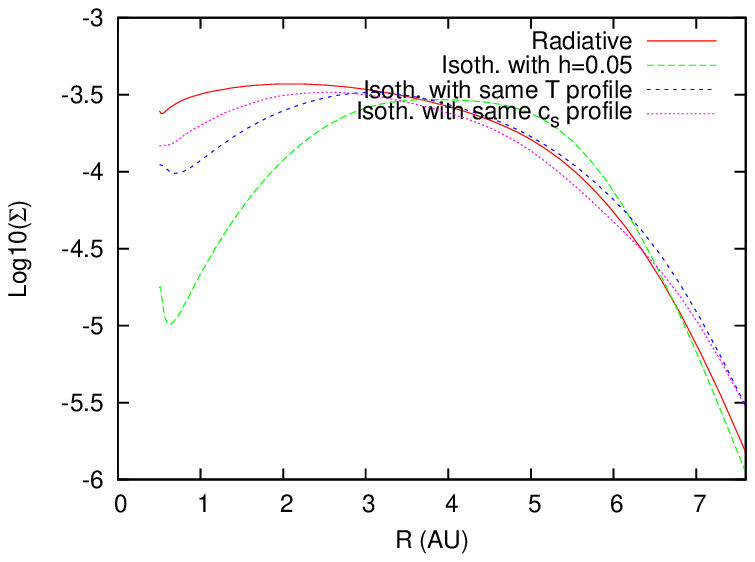}
   } 
   \centering\resizebox{90mm}{!}  {
     \includegraphics{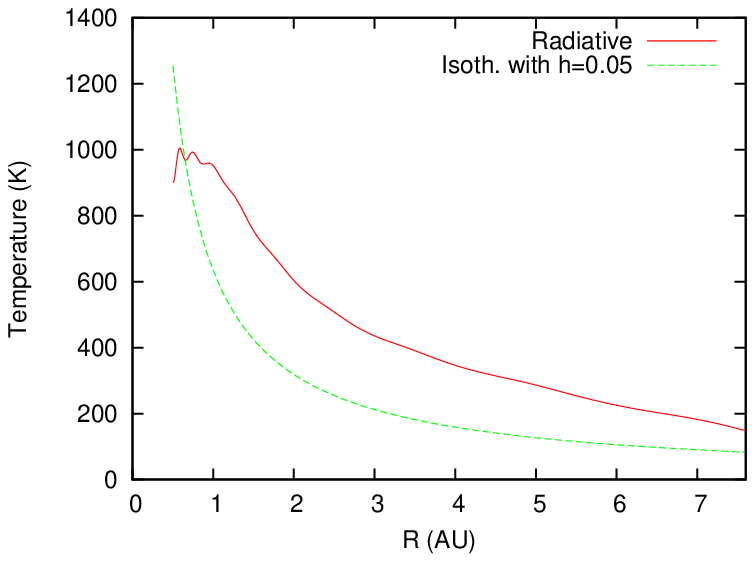}
   }
   \caption{\label{f_irr}Time- and azimuthally-
     averaged disc eccentricity, density and temperature
     profiles for our radiative disc model and various locally isothermal 
     disc models: (i) one with the same initial aspect ratio as in the radiative 
     model ($h = 0.05$), (ii) one with the same temperature 
     profile as in the radiative model, and (iii) another with 
     same sound speed profile as in the radiative model.
     }
\end{figure*}
%FFFFFFFFFFFFFFFFF
%\begin{figure}
%  \includegraphics[width=\hsize]{temp.eps}
%  \caption{\label{f_temp}Temperature profiles of discs with 3
%    different energy equations: isothermal (red line), radiative
%    (green line) and irradiated (blue line).  The binary orbital
%    parameters are $a_b = 30$ AU, $e_b=0.4$.{\color{blue}We should now show the results for the locally isothermal run with prescribed temperature profile.}}
%\end{figure}
%FFFFFFFFFFFFFFFFF
%FFFFFFFFFFFFFFFFF
%FFFFFFFFFFFFFFFFF
%\begin{figure}
%  \includegraphics[width=\hsize]{f_prof.eps}
%  \caption{\label{f_prof}Averaged disc eccentricity radial dependence
%    for three different models: radiative, isothermal with the same
%    temperature profile of the radiative one, isothermal with the same
%    sound speed profile of the radiative one. The disc eccentricity is
%    by far larger in the isothermal models.}
%\end{figure}
%FFFFFFFFFFFFFFFFF
As a first step, we compare a radiative and a locally isothermal model
with same initial aspect ratio, $h=0.05$. All other disc parameters are 
as described in Sect.~\ref{sec:model}. Fig.~\ref{f_2D} shows contours 
of the disc's surface density at 90 binary orbits obtained with two values 
of the binary's eccentricity: $e_b=0.4$ (upper panels) and $e_b=0$ (lower 
panels). As already pointed out in Sect.~\ref{sec:isoth}, eccentric 
streamlines ($m=1$ density mode) have different values of the pericenter depending on the
radial distance, and, as a consequence, they combine into a pattern of
spiral structure. However, the discs computed with the radiative model
(left-handed plots in Fig.~\ref{f_2D}) appear smoother and more symmetric than the
corresponding locally isothermal discs, and, in particular, they do not
feature any elliptic hole near the inner edge, independently of $e_b$. The
spiral density waves are less strong in the radiative case and this is 
possibly related to the absence of a low density region close to the star. 
The disc eccentricity $e_d$ and perihelion longitude are shown as a function of time in
Fig.~\ref{f_eperi} and they confirm that the radiative case has, on
average, a lower eccentricity even if it takes more time 
to reach a steady state.  The perihelion libration, observed also for
the radiative case, strongly suggests that this behaviour is solely
due to the disc self-gravity and that it does not depend on the energy
equation.  By inspecting the radial pericenter profile, even in the
radiative case the azimuthally averaged pericenter of the gas
streamlines changes with radial distance (as is illustrated in 
Fig~\ref{f_ome} for a locally isothermal model). This variation, in addition to 
causing spiral waves, leads to an asymmetric distribution of mass and
then to a non-symmetric disc gravity field. This is critical for
planetesimals embedded in the disc since the non-homogeneous disc
forces non-radial components on the gravity field felt by
planetesimals significantly perturbing their orbits.  These
perturbations are comparable in magnitude to the secular effects of
the companion star but are irregular and may then cause larger changes
in the planetesimals orbital elements. They are indeed an indirect
effect of the companion gravity but for planetesimals they represent
an independent source of perturbation.

\begin{figure*}[hpt]
  \centering\resizebox{\hsize}{!}  {
    \includegraphics{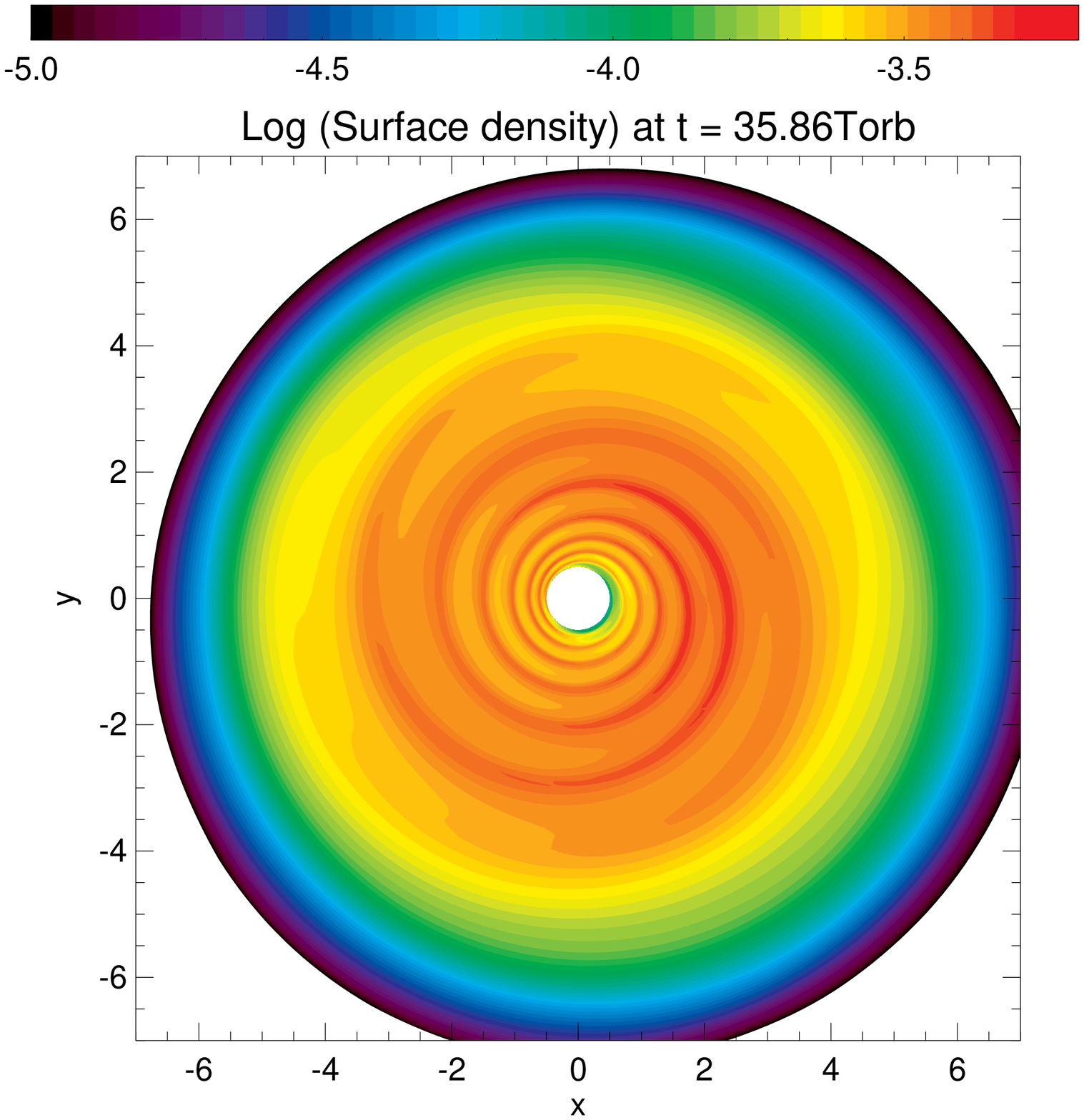}
    \includegraphics{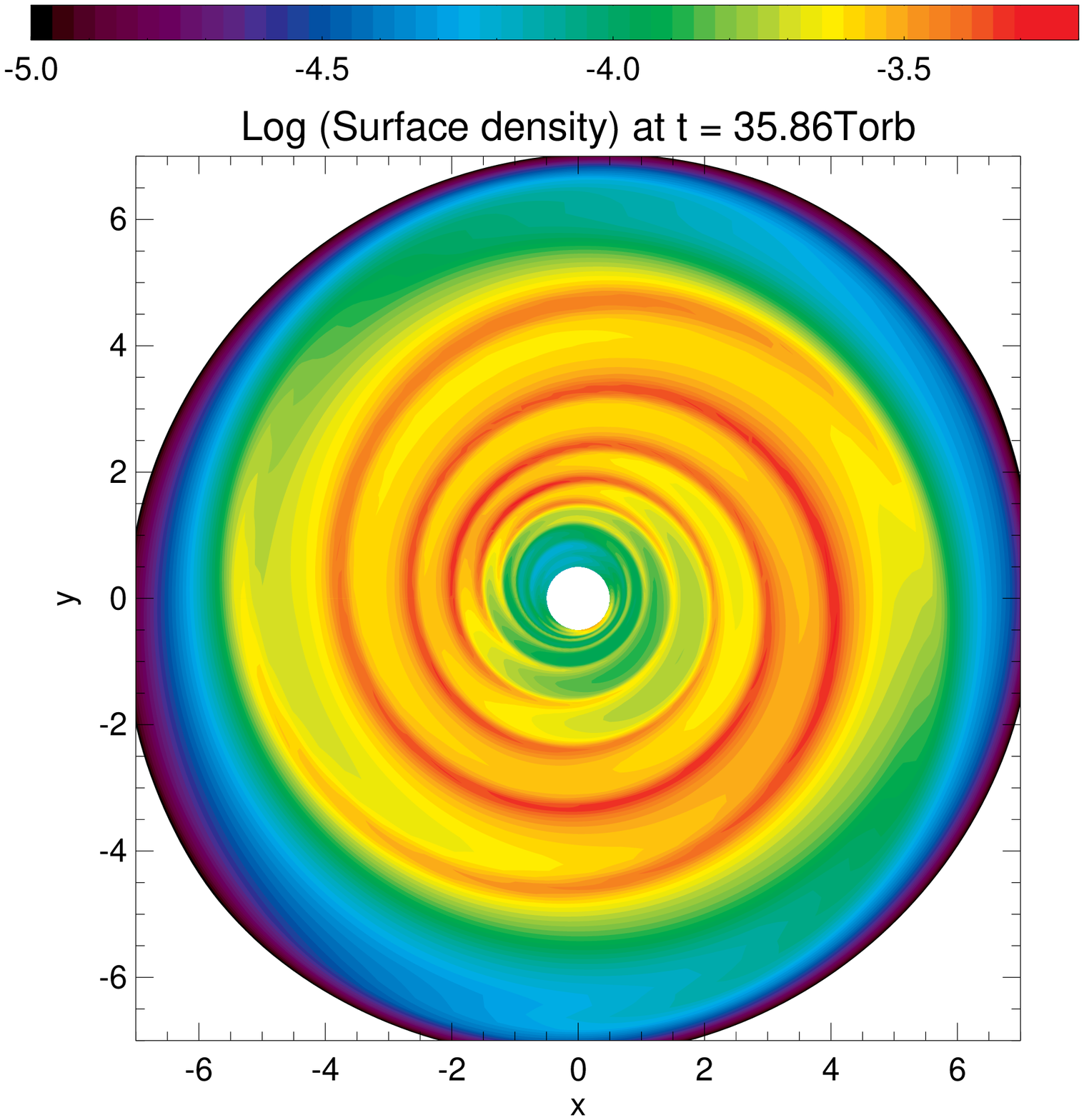}
   } 
   \caption{\label{comparison}Contours of the disc surface density for the
     radiative model (left panel), and the locally isothermal model with same
     imposed temperature profile as in the radiative case (right panel).  In the
     radiative case, spiral waves appear smoother and more
     damped.}
\end{figure*}

To get further insight into the different densities and eccentricities  
with and without an energy equation, we carried out two additional 
locally isothermal disc models. In the first model, the initial temperature 
is set to the time-averaged temperature profile of the above radiative 
run. In the second model, the initial temperature is chosen to give 
the same sound speed profile as the time-averaged one in the 
radiative run. In Sect.~\ref{sec:31}, we pointed out that the disc eccentricity in locally
isothermal models strongly depends on the aspect ratio $h$ and, as a
consequence, on the temperature profile (see Fig.~\ref{f_iso}).  In
Fig.~\ref{f_irr}, bottom plot, we compare the temperature profile of
the locally isothermal model with $h=0.05$ to  the time-averaged 
temperature profile of the radiative  model. 
The temperature in the radiative run is significantly larger, 
implying that we should compare the radiative model to a locally
isothermal model with $h \sim 0.07$. In the top panel of Fig.~\ref{f_irr}, 
we display the eccentricity and density profiles in
all four models: (i) the radiative model, (ii) the locally isothermal model 
with same initial aspect ratio as in the radiative model, (iii) the locally isothermal model 
with same temperature profile as in the radiative case, and (iv) the locally 
isothermal model with same sound speed profile as in the radiative one. 
The disc eccentricity in models (iii) and (iv) is smaller than in model (ii), 
 as expected from the results shown in Fig.~\ref{f_iso} (top-left panel). 
 Consequently, the disc surface density remains larger in models (iii) and (iv) 
 compared to in model (ii). Still, models (iii) and (iv) do not match the low eccentricity of the
radiative case and its smooth density distribution close to the
star. The strength of the $m=1$ Fourier mode in the radiative run is
smaller in the disc's inner parts, and this difference depends on
the form of the energy equation. In Fig.~\ref{comparison}, we display 
density contours obtained with the radiative run (left panel), and with the locally isothermal 
model with same temperature profile (right panel). In the radiative model, the wave
perturbations are less strong
and the waves appear to be more damped. Being
weaker in the inner disc parts in the radiative case, waves should be 
less efficient at depositing angular momentum there. This would explain why
radiative discs are less eccentric in the regions close to the central
star, and why the surface density is more homogeneous.

%FFFFFFFFFFFFFFFFF
%FFFFFFFFFFFFFFFFF
% ---------------------------------------
\subsection{Radiative damping of waves}
% ---------------------------------------

The main question arising from previous results is what causes 
a stronger damping of density waves in the radiative model than in the 
locally isothermal one, while the same (time-averaged) sound speed 
profile is adopted in both models. In Fig.~\ref{f_irr}, 
the radiative model shows indeed a significantly lower eccentricity profile.
Assuming the disc eccentricity is related to the strength of the 
density waves induced by the binary gravitational perturbations, 
we have shown in Fig.~\ref{comparison} that spiral waves are more 
damped in the radiative case.

There are two potential sources of wave damping that may 
help understand why the disc eccentricity remains lower in the radiative case: 
shock damping \citep{gora} and radiative damping \citep{cassen}. To explore the efficiency of 
the shock damping mechanism, we examined the vortensity distribution 
in the disc, since it experiences a jump at shocks, the magnitude of which depends on
the strength of the shock. We did not observe significant differences 
in the disc vortensity distribution between the two models and thus we 
believe that 
shock damping does not significantly contribute to 
the smoother behaviour of the radiative disc.

The energy loss by radiation is an additional possible source of 
wave damping \citep{cassen}. Wave propagation through adiabatic 
compressions and expansions may be damped 
by radiative losses which, in our model, are controlled by 
the cooling term $Q^-_{cool}$. To test this hypothesis, we restarted 
our standard radiative simulation ($e_b=0.4$,
$M_{\rm s} = 0.4M_{\odot}$ and $a_{\rm b} = 30$ AU) after 165 binary 
revolutions adopting different two cooling prescriptions. In a first 
run, we limit the cooling time throughout the disc to be no 
less than that at 6 AU from the star (that is, about 150 yr). To keep the temperature 
profile as close as possible to that of the standard radiative model,
and thus to prevent discrepancies related to different temperature profiles,  
we limit the viscous heating timescale by a similar amount. In a second run,  
we increase the cooling rate throughout the entire disc by a factor of 10, 
and the viscous heating rate is increased accordingly. 
Both restart simulations were run for 50 additional binary orbits, over which 
the disc's temperature profile does not evolve significantly, except within 1 AU 
from the central star. 

\begin{figure}[hpt]
    \includegraphics[width=\hsize]{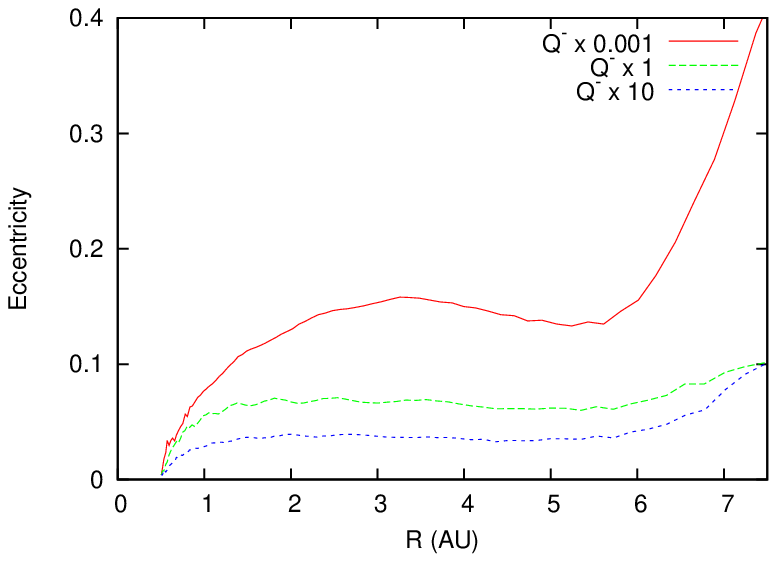}
    \includegraphics[width=\hsize]{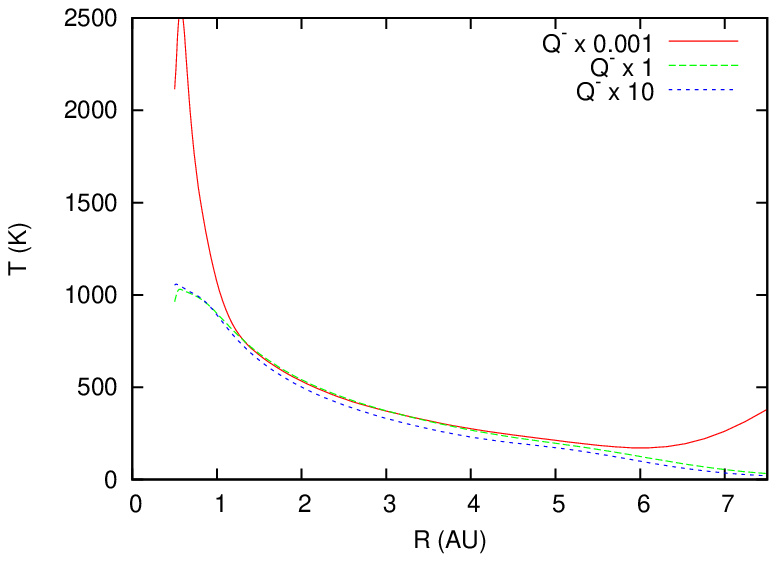}
  \caption{\label{damping} \bf Disc eccentricity and temperature
profiles (azimuthally- and time-averaged) for our standard radiative 
model (labelled as $Q^- \times 1$), and for two restart simulations 
with different cooling rates. The disc eccentricity is
higher in the restart run with smaller cooling rate ($10^{-3} \times Q^-$), while it is lower 
in the restart simulation with larger cooling rate ($10 \times Q^-$).}
\end{figure}

Fig.~\ref{damping} compares the disc time-averaged eccentricity and temperature
profiles of the standard radiative model with those of the additional models with 
longer and shorter cooling times. The inner disc's eccentricity is significantly increased 
with longer cooling timescale, exceeding $\sim 0.1$ almost uniformly 
from $R=2$ AU to $R=6$ AU. This value is in good agreement with that of the
locally isothermal run with same sound speed profile (see Fig.~\ref{f_irr}). 
However, close to the inner edge the disc eccentricity is still small 
but this is possibly due to the fact that we cannot maintain constant the temperature
profile within 1 AU of the central star in the model with reduced cooling. 
In the restart simulation with a cooling rate 10 times larger, the disc
eccentricity is about half that of the standard radiative model, while the corresponding 
temperature profiles hardly differ.

The results shown in Fig.~\ref{damping} suggest that radiative damping 
is responsible for the limited growth in the disc eccentricity compared to  
to similar locally isothermal disc models. This mechanism, like self-gravity, 
helps maintaining the disc eccentricity to a mild value.

%FFFFFFFFFFFFFFFFF
%FFFFFFFFFFFFFFFFF

% ---------------------------------------
\subsection{Dependence on the boundary conditions}
% ---------------------------------------
\label{sec:boundary}
We examine in this paragraph the dependence of our results 
on the choice for the boundary condition at the grid's inner edge. 
For this purpose, we compare the results of simulations using 
our fiducial boundary condition (zero-gradient outflow boundary condition, see 
Sect.~\ref{sec:rde}) with those of two additional simulations:
\begin{enumerate}
\item[-] One using a viscous outflow boundary condition \citep{kn08}. It 
is very similar to our zero-gradient outflow boundary 
condition, except that the radial velocity at the inner boundary is set 
to the local (azimuthally-averaged) viscous inflow velocity of a disc 
in equilibrium with locally uniform surface density profile ($-3\nu / 2R_{\rm min}$).
The azimuthal velocity is also set to its initial axisymmetric value,
\item[-] Another simulation also using our standard outflow boundary 
condition, but where the azimuthal velocity at the inner boundary 
is extrapolated from that in the first active ring with a $r^{-1/2}$ 
law. In contrast to previous boundary conditions, the disc at the grid's inner 
edge is no longer forced to remain circular.
\end{enumerate}

\begin{figure*}[hpt]
  \centering\resizebox{\hsize}{!}  {
    \includegraphics[width=0.32\hsize]{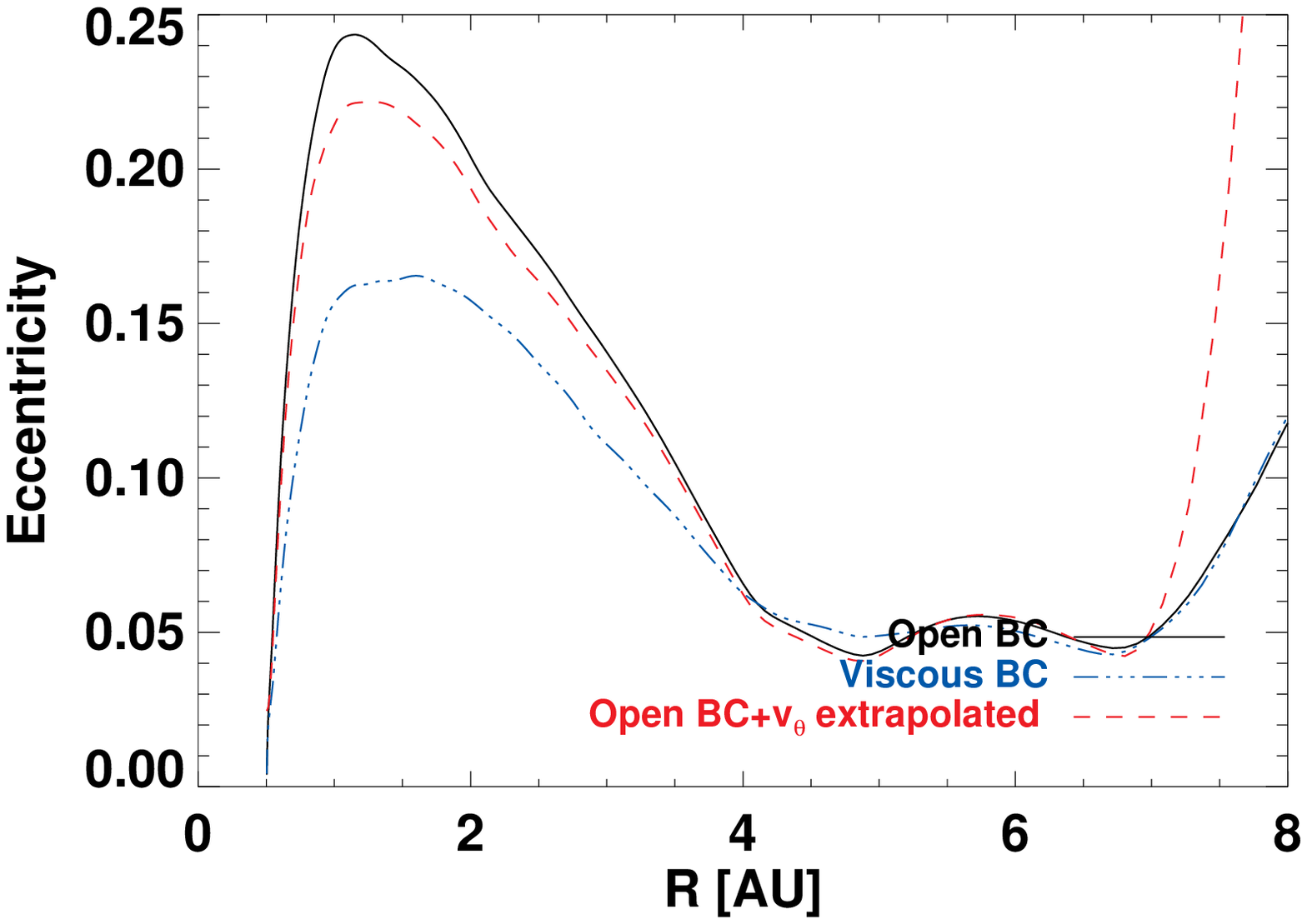}
    \includegraphics[width=0.32\hsize]{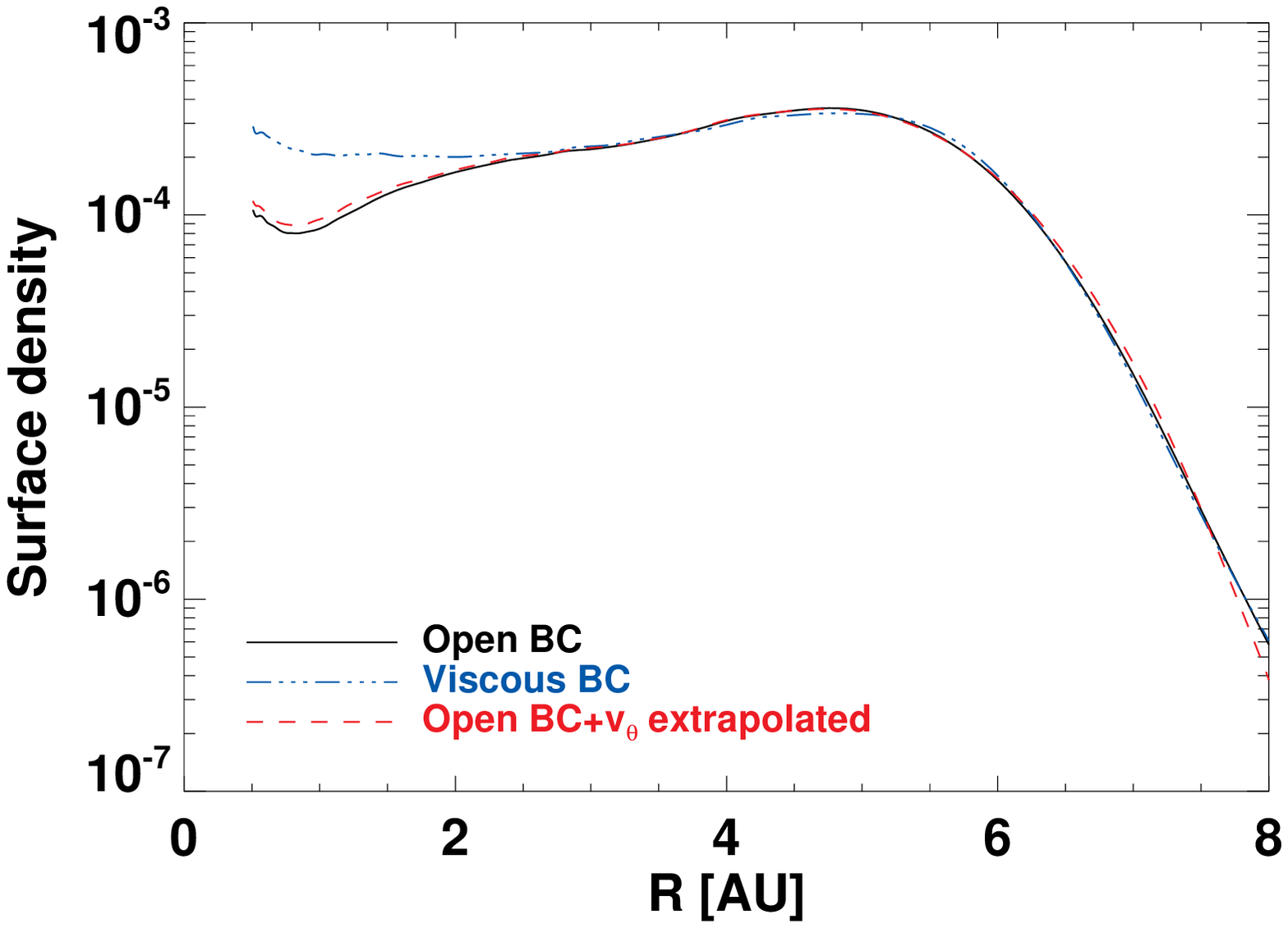}
   }
   \caption{\label{boundary_iso} Eccentricity and surface density profiles, time-averaged 
   between 35 and 40 binary orbits, obtained with the locally isothermal disc model 
   with imposed temperature profile, using three 
   different boundary conditions at the inner edge: (i) our standard zero-gradient outflow 
   boundary (labelled as open, solid curve), (ii) a viscous outflow boundary (labelled 
   as viscous, dash-dotted curve, see text), and (iii) our standard zero-gradient outflow 
   boundary with the azimuthal velocity at the inner edge extrapolated from that in the 
   first active ring (dashed curve).}
\end{figure*}

\begin{figure*}[hpt]
  \centering\resizebox{\hsize}{!}  {
    \includegraphics[width=0.32\hsize]{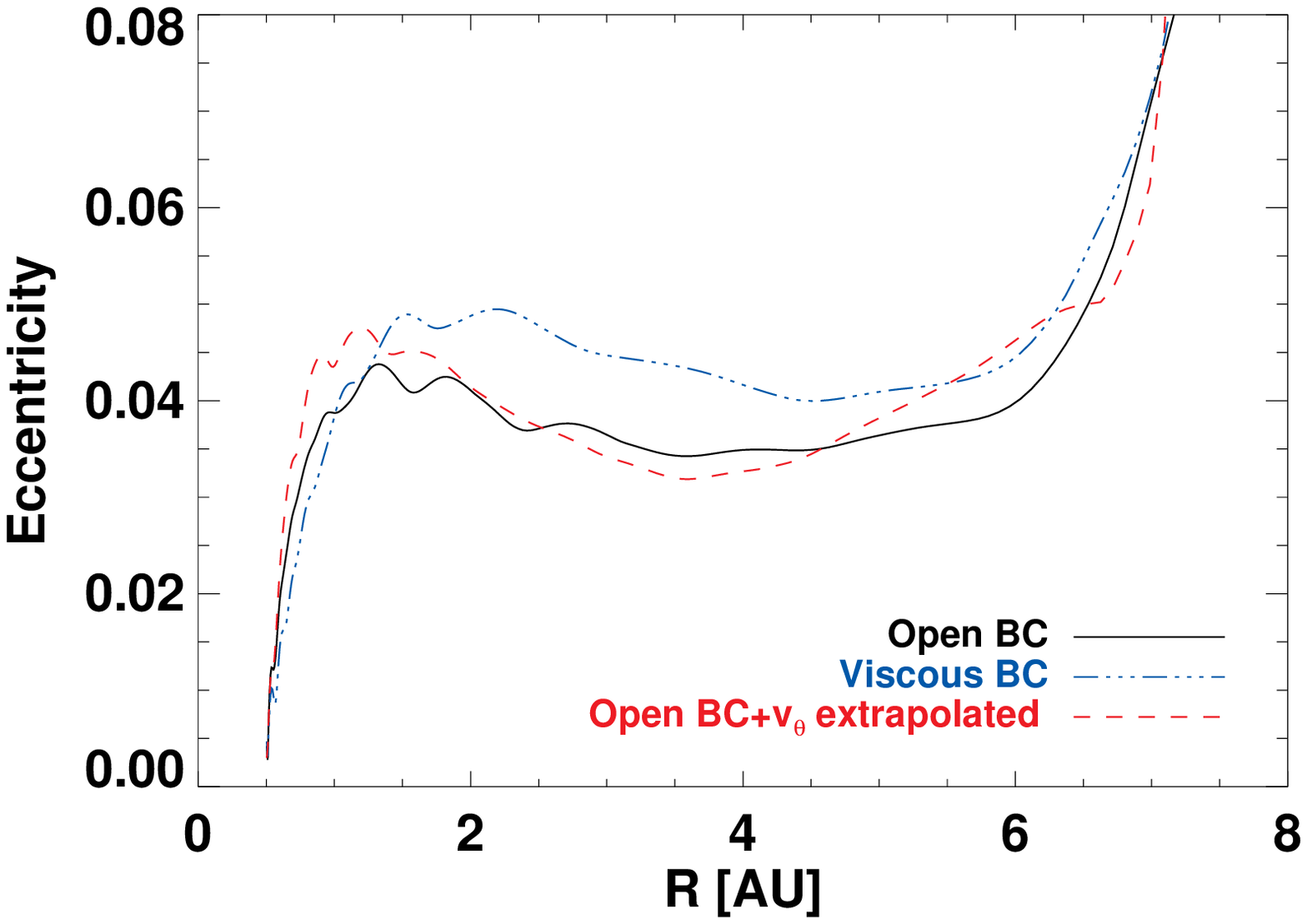}
    \includegraphics[width=0.32\hsize]{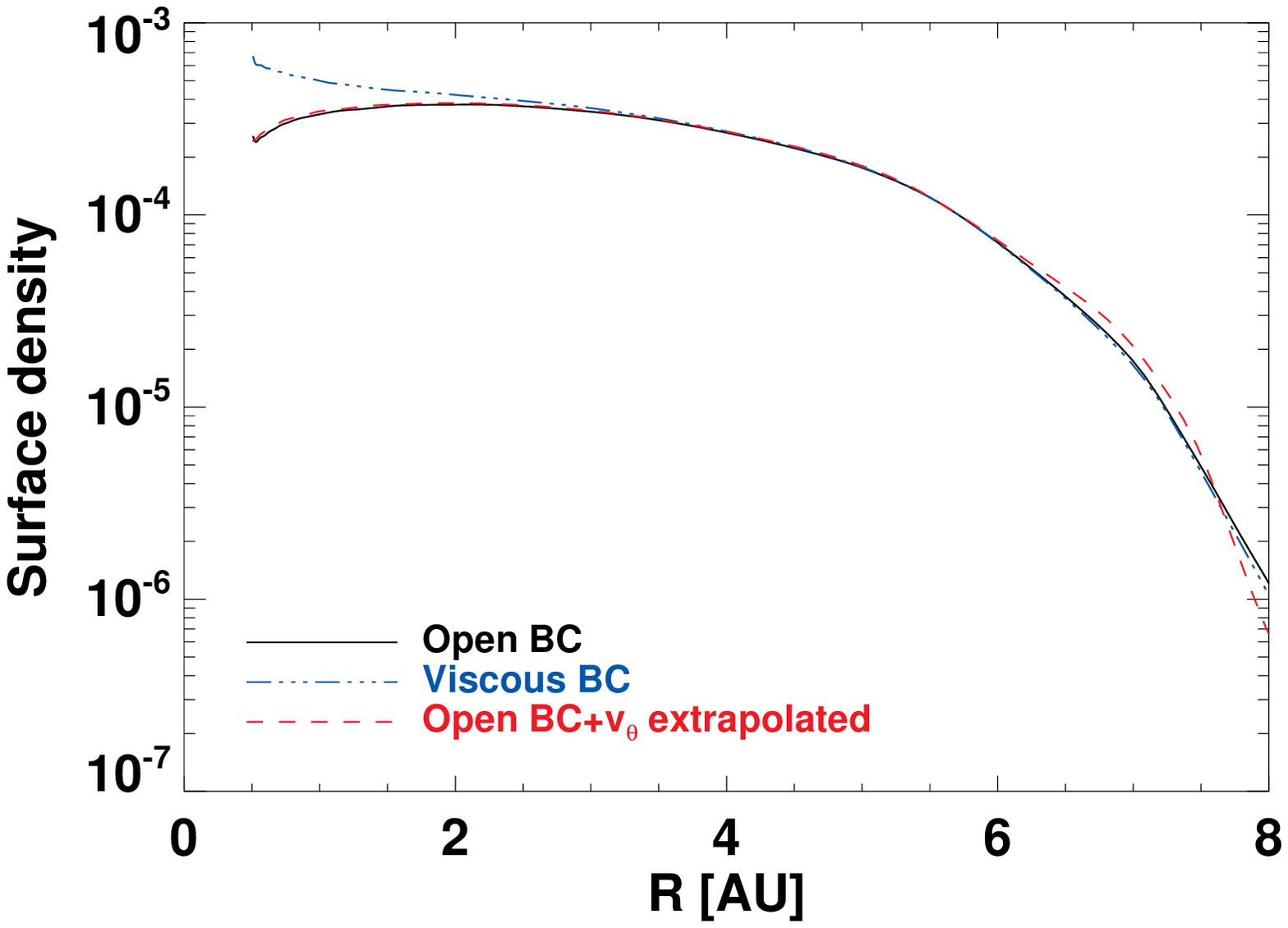}
   }
   \caption{\label{boundary}  Same as in Fig.~\ref{boundary_iso}, 
   but for our standard radiative disc model.}
\end{figure*}

The results of this comparison for locally isothermal runs with same temperature 
profile as in the radiative model are shown in Fig.~\ref{boundary_iso}.
The profiles are in good agreement for all three different boundary conditions. 
Note that the surface density in the inner disc for the viscous boundary condition is 
larger than with the two other boundaries. This is expected, since the imposed viscous drift velocity is smaller 
that the radial velocity set by the propagation of the spiral waves induced by the secondary. 
A similar good agreement is obtained with a radiative model, as illustrated in 
Fig.~\ref{boundary}. These results confirm the robustness of our 
results against the choice for the inner boundary condition.

%FFFFFFFFFFFFFFFFF
%FFFFFFFFFFFFFFFFF
%FFFFFFFFFFFFFFFFF
%FFFFFFFFFFFFFFFFF
% ---------------------------------------
\subsection{Dependence on the binary eccentricity}
\begin{figure*}[hpt]
  \centering\resizebox{\hsize}{!}  {
    \includegraphics[angle=-90,width=0.32\hsize]{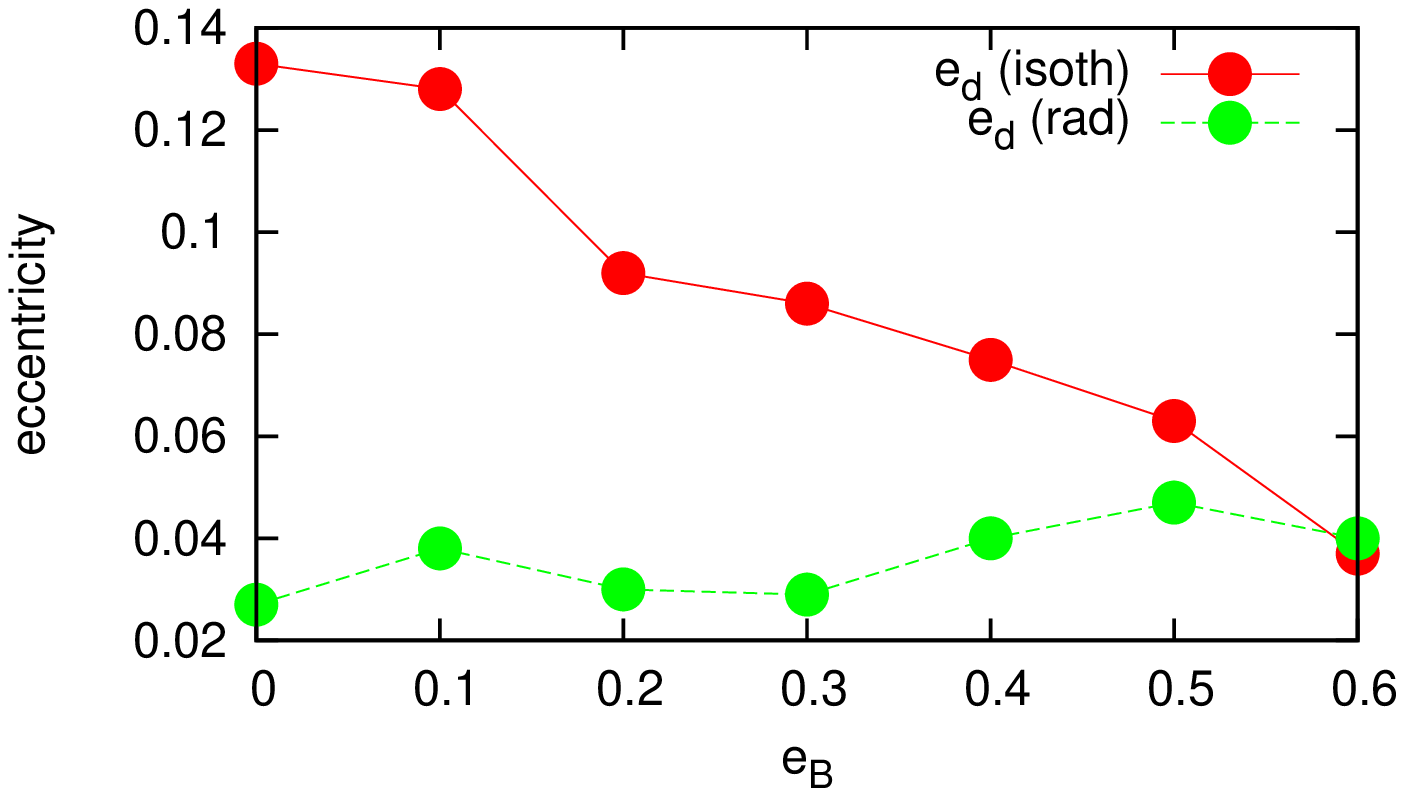}
    \includegraphics[angle=-90,width=0.32\hsize]{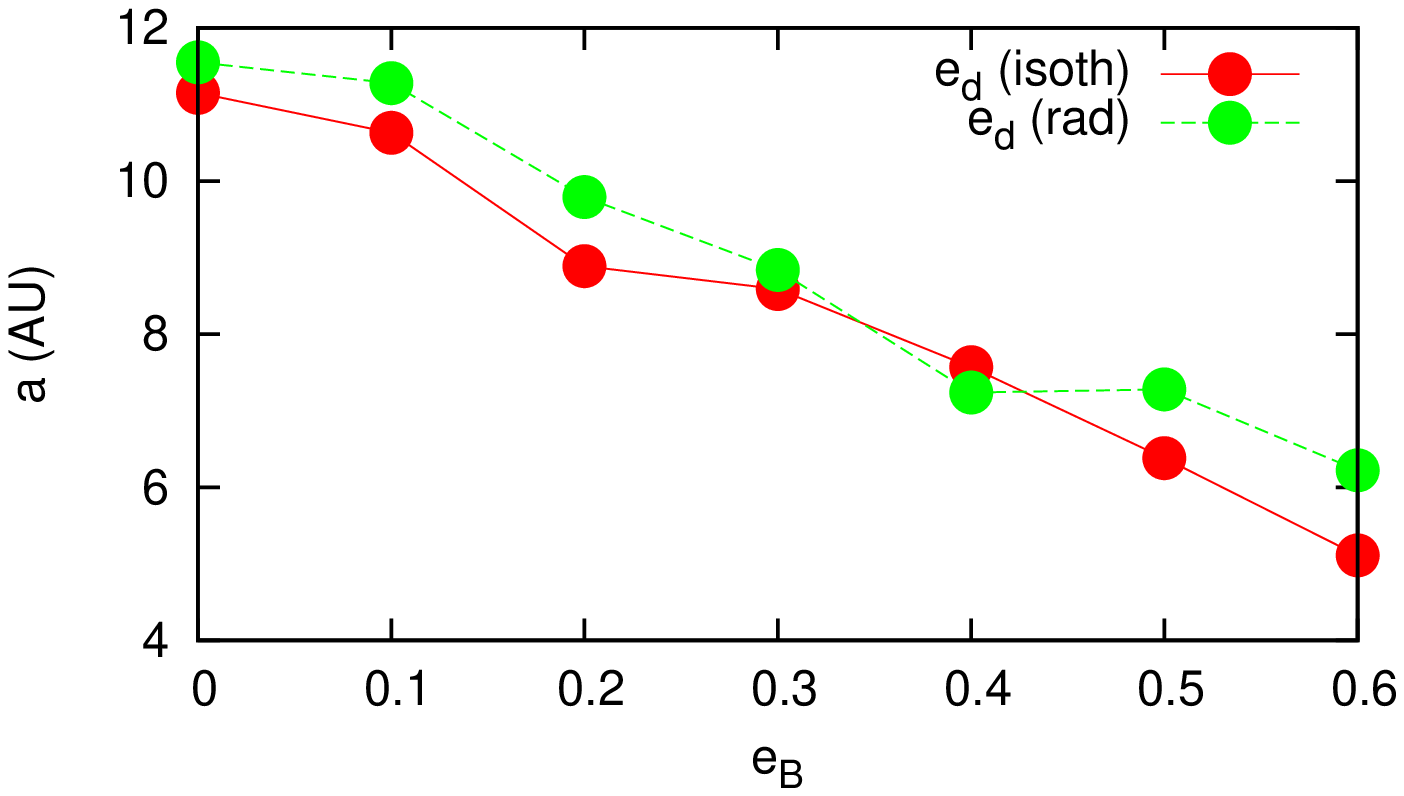}
   }
   \caption{\label{f_e_a}Disc eccentricity (left plot) for different
     values of the binary eccentricity $e_b$. The average
     of $e_d$ is displayed after 70 binary revolutions, and time-averaging 
     is done over 20 revolutions. For comparison, we also depict the 
     corresponding values of $e_d$
     for isothermal discs with same binary configurations
     \citep{mabash}. In the right plot, we show the disc's size 
     for different values of $e_b$, more specifically the semi-major
     axis of the ellipse that best fits the disc density at a level 
     $\Sigma = 10^{-6}$ (code units). }
\end{figure*}

\begin{figure}[hpt]
  \includegraphics[width=\hsize]{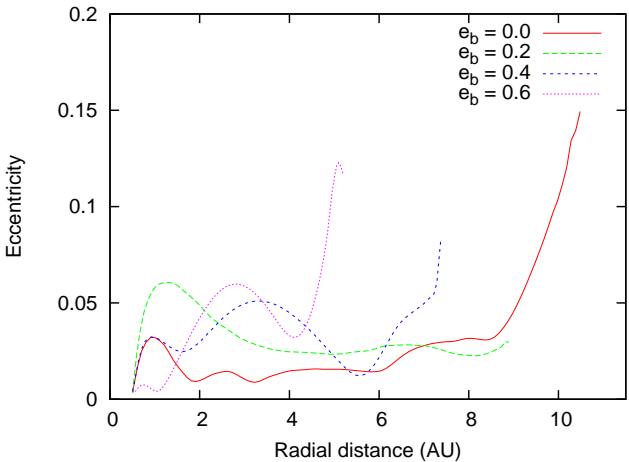}
  \caption{\label{f_25_eb}Azimuthally-averaged eccentricity profiles for 
    radiative discs with different binary eccentricities, displayed 
    at 100 binary orbits. Instantaneous profiles are depicted to better 
    highlight the dependence of $e_d$ with $e_b$. }
\end{figure}

% ---------------------------------------
In paper I, we showed that the disc eccentricity $e_d$  decreases 
with increasing binary eccentricity, the  circular case ($e_b=0$) being
the most perturbing configuration. We interpreted this result as being
due to the larger size of the disc for lower values of $e_b$, and to
the consequent larger number of resonant perturbations which may
affect it. The eccentricity of radiative discs, however, seems to be 
rather insensitive to $e_b$. The outcome of the simulations
at different $e_b$ is shown in the left plot of
Fig.~\ref{f_e_a}, where we compare the values of $e_d$ obtained for
locally isothermal and radiative discs. In contrast to locally
isothermal discs, the averaged $e_d$ of radiative discs is
almost constant around 0.05, and it does not show a significant
dependence with $e_b$. However, the azimuthally-averaged profile 
of the disc's eccentricity over the radiative disc does 
depend on $e_b$. In Fig.~\ref{f_25_eb}, the radial profile of the disc's 
eccentricity is shown for different values of $e_b$ ranging from 0 to
0.6. We note that the eccentricity profiles significantly differ even
if the position of the star with respect to the initial reference
frame is the same. All radiative runs show an increasing eccentricity
profile towards large radii, whereas locally isothermal runs feature a peak 
in the eccentricity in the inner disc parts for most values of $h$.
It should be pointed out that the above 
comparison is done by adopting the same initial aspect ratio for radiative 
and locally isothermal disc models. A different approach would 
be to compare disc models with same temperature or sound speed profiles, 
as is done in Figs.~\ref{comparison} and~\ref{f_irr}. But, when comparing 
the temperature profiles of the radiative discs, we find that they correspond 
to locally isothermal discs with $h \sim 0.06-0.07$, which anyway feature 
a larger eccentricity in the disc inner parts.

%FFFFFFFFFFFFFFFFF
%FFFFFFFFFFFFFFFFF
%FFFFFFFFFFFFFFFFF
%FFFFFFFFFFFFFFFFF
%FFFFFFFFFFFFFFFFF
\begin{figure}[hpt]
  \centering\resizebox{\hsize}{!}  {
    \includegraphics[angle=-90,width=0.35\hsize]{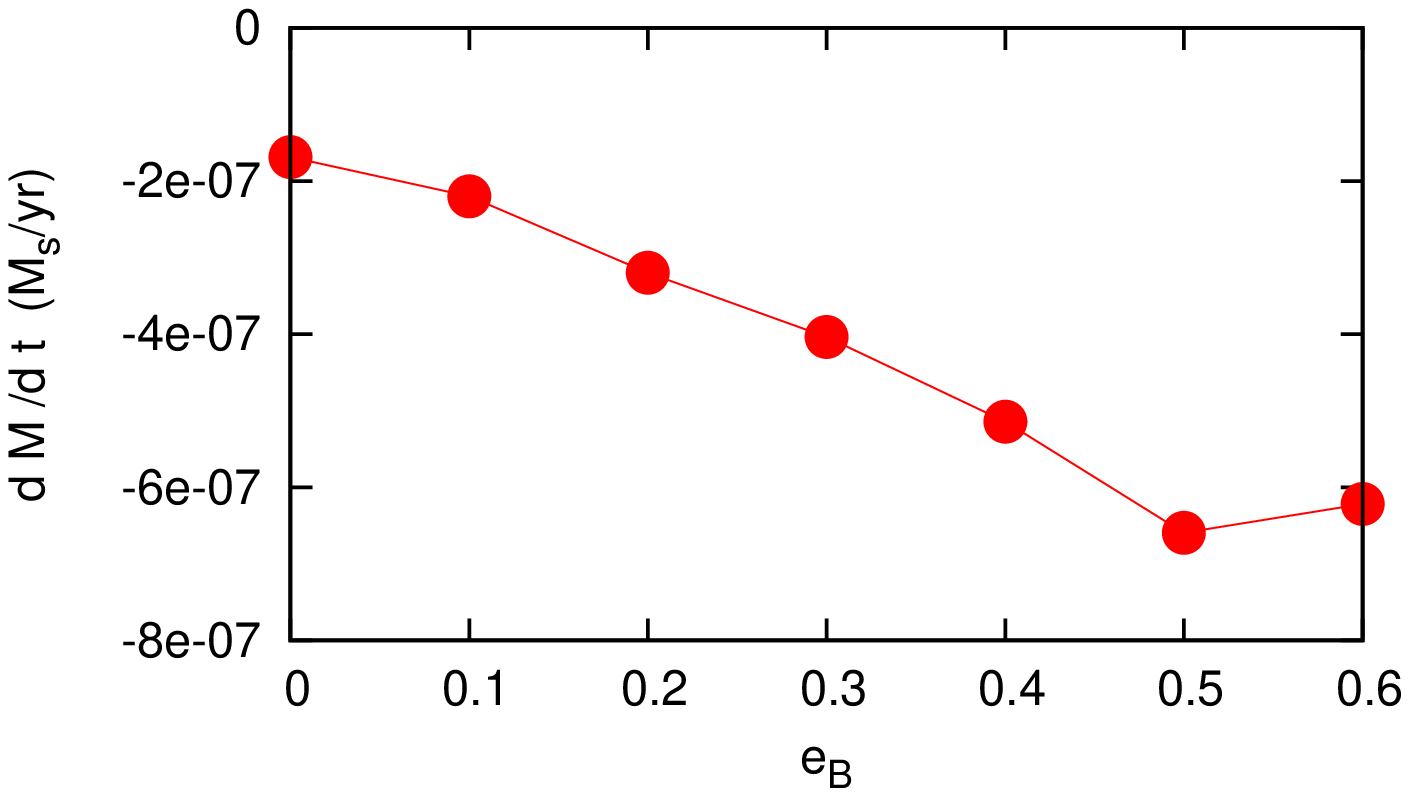}
   }
   \caption{\label{f_mass_loss}
     Mass loss rate of radiative discs (in $M_{\odot}/yr$) for different
     values of $e_b$ 
     obtained through a linear fit of the time evolution of the 
     disc mass after the initial fast truncation.}
\end{figure}
%FFFFFFFFFFFFFFFFF

In  the right panel of Fig.~\ref{f_e_a} we also compare the values of the outer semi-major
axis $a_d$ of the ellipse best fitting the outer edge of the disc (at
a density level of $\Sigma = 10^{-6}$, code units).  The radiative discs extend
slightly farther out compared to the corresponding locally isothermal discs, 
except for our fiducial binary's eccentricity ($e_b=0.4$), where the disc's 
size is approximately equal to its isothermal counterpart. The mass loss rate
for radiative discs, for different values of $e_b$, is shown in 
Fig.~\ref{f_mass_loss}. It is computed with a
linear fit of the time evolution of the disc mass when a steady state is reached,
so it is independent from the initial disc truncation process. 
It shows an almost linear dependence with $e_b$  and it is related to 
the strong perturbation effects on the disc as the secondary star passes 
at pericenter. The inspection of the amount of mass stripped vs. time 
reveals that almost all the mass loss occurs 
during and after the 
pericenter passage of the companion while a lower amount is 
constantly lost  
through the inner border due to the disc eccentricity and viscosity. 
The absolute value of the mass loss is large, and it predicts a reduction in the
disc mass by a factor of two in about $3 \times 10^4$ yrs for our
standard case ($e_b = 0.4$). However, this value depends on the disc
mass, and a simulation with the same binary parameters but an initial
disc mass 10 times smaller gives a mass loss rate $\approx 3.7
\times 10^{-9}$ $M_{\odot}/yr$. 

% ---------------------------------------
\subsection{Dependence on the disc mass}
% ---------------------------------------

%FFFFFFFFFFFFFFFFF
\begin{figure}[hpt]
  \includegraphics[width=\hsize]{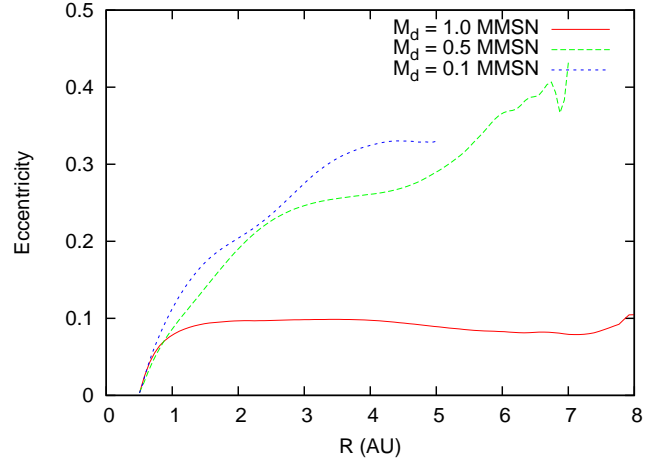}
  \caption{\label{f_mass}Azimuthally- and time-averaged profiles 
  of the disc's eccentricity 
%(upper panel) and temperature 
%  (lower panel) 
for three different initial surface densities: 
  $\Sigma_0 = \Sigma_{\rm MMSN}$ (our fiducial value), $\Sigma_0 = 0.5
    \Sigma_{\rm MMSN}$, and $\Sigma_0 = 0.1 \Sigma_{\rm
      MMSN}$. }
\end{figure}
%FFFFFFFFFFFFFFFFF

\begin{figure}[hpt]
  \includegraphics[width=\hsize]{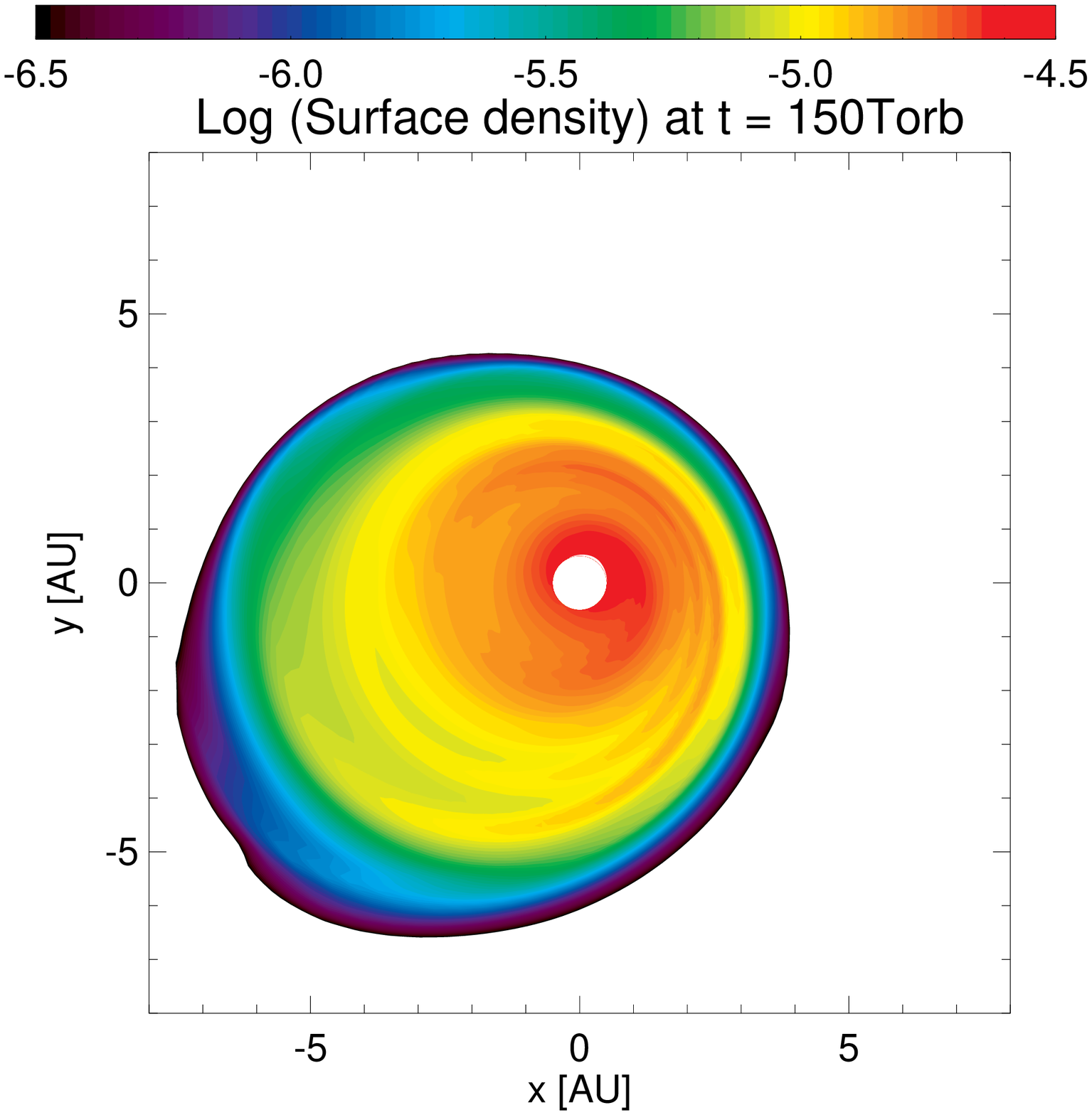}
  \caption{\label{f_mass_2D}Contours of the gas density after 150
    binary revolutions for a radiative disc with $\Sigma_0 = 0.1
    \Sigma_{\rm MMSN}$.  }
\end{figure}

In Paper I, we found that for locally isothermal discs, a
decrease in the disc mass had no significant impact on the disc
eccentricity. Below we show that radiative discs behave differently.  In
Fig.~\ref{f_mass}, we compare the disc eccentricity in the nominal case,
where the initial density at 1 AU is that of the MMSN, to that of a
disc initially 2 and 10 times less massive.  The disc eccentricity
$e_d$ in a steady state is much larger for the less massive discs. This
is further illustrated by the contours of the disc density for $\Sigma_0 = 0.1
\Sigma_{\rm MMSN}$ in Fig.~\ref{f_mass_2D}. The disc appears
smaller and very eccentric, particularly in its outer regions. 

%FFFFFFFFFFFFFFFFF
%FFFFFFFFFFFFFFFFF

\begin{figure}[hpt]
  \includegraphics[width=0.7\hsize,angle=-90]{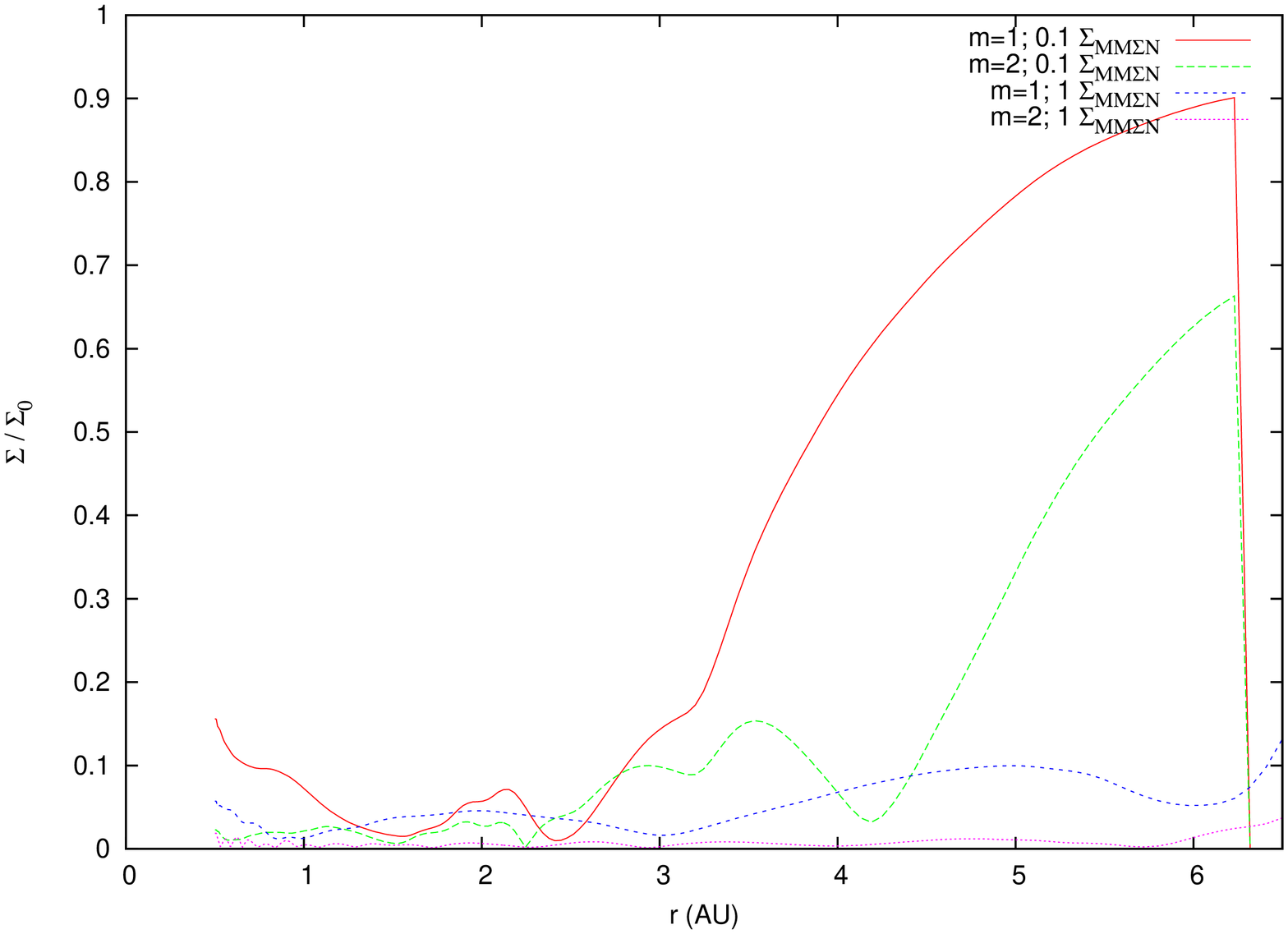}
  \caption{\label{f_foum}$m=1$ and $m=2$ Fourier components of the
    normalized surface density for a radiative disc with $\Sigma_0 = 
    \Sigma_{\rm MMSN}$, and a radiative disc initially ten times less 
    massive ($\Sigma_0 = 0.1\Sigma_{\rm MMSN}$).}
\end{figure}

It is difficult to 
interpret this outcome on the basis of the isothermal
simulations. Less massive radiative discs, like that shown 
in Fig.~\ref{f_mass_2D}, have lower temperature
profiles. This is presumably due to a faster cooling rate. According 
to our model, a lower surface density of the disc implies a 
smaller optical thickness that leads to a shorter 
cooling timescale. As a consequence, our standard model with $\Sigma_0 = 1
\Sigma_{\rm MMSN}$  has a 
stationary temperature profile which is equivalent to
that of a locally isothermal run with $h \sim 7-8\%$ while
less massive discs have
a lower temperature profile in a steady-state with a 
temperature typically $3-4$ times less high. 
This translates into equivalent
locally isothermal models with $h$ down to $3-4\% $. By inspecting  Fig.~\ref{f_iso} 
(top left plot)
we notice that 
in this range of aspect ratios, the disc's eccentricity actually
changes quickly with $h$. This might explain why less massive
radiative discs have different disk eccentricity. Even self--gravity 
will play a minor role for less massive discs and in Paper I we 
showed that indeed self--gravity is effective in reducing the 
disc eccentricity. This, however, is not the full story and 
the different energy equation also plays a significant role. 
The disc eccentricity radial profile of radiative discs, also the less massive 
ones (see Fig.~\ref{f_mass}), does not peak close to the 
star like for the isothermal discs (Fig.~\ref{f_iso}). The value 
of $e_d$ for radiative discs grows
for larger values of $R$ as predicted by the
analytical theory of \cite{paard08}.
This behaviour is further 
illustrated by examining the Fourier components of the normalized
surface density. Fig.~\ref{f_foum} compares the Fourier
components for both our standard disc with $\Sigma_0 = \Sigma_{\rm
  MMSN}$, and the disc model with $\Sigma_0 = 0.1 \Sigma_{\rm MMSN}$. The
$m=1$ and $m=2$ components are much stronger for the less massive disc,
and they account for the overall higher disc eccentricity. In the
standard case, the high disc density is able to damp efficiently the
two Fourier components of the binary perturbations as they move toward 
the disc inner parts. On the other hand, these components propagate further 
in for less massive discs, leading to a larger value for $e_d$. However, also in the
less massive disc they do not reach the inner part of the disc, like in 
isothermal discs, preventing the formation of an eccentric 
low-density region close to the central star. 

%FFFFFFFFFFFFFFFFF
%FFFFFFFFFFFFFFFFF

According to the results presented so far, isothermal and radiative
discs behave differently. Massive isothermal discs in low eccentricity
binaries are expected to be eccentric for reasonable values of $h$
while radiative discs always have low eccentricity.  For isothermal
discs there is a weak dependence of the disc eccentricity on the disc
mass (see Paper I) while, for radiative discs, this dependence is
strong with less massive discs being more eccentric.  Hot isothermal
discs develop an inner eccentric hole while radiative discs are smooth
also at the inner edge.

% ====================== %
\section{Why is disc eccentricity so important? Planetesimals!}
% ====================== %
The shape and profile of the gas disc may have a crucial role in the
early stages of planetary formation, when kilometer-sized
planetesimals are colliding with each other. Even when neglecting gas
disc gravity, several studies have shown that the coupling between
secular perturbations and gas drag, which increases impact velocities
between non-equal sized bodies and can lead to accretion-hostile
environments \citep[e.g.][]{theb06, theb08, theb09}, leads to further
increasing the relative velocity between planetesimals if the gas disc
gets eccentric \citep{paard08,xie10}.  Disc gravity might further
increase impact velocities by increasing the disc eccentricity and
inducing additional dynamical perturbations on planetesimals because
of non-isotropic distribution of the mass within the gas disc.  The
asymmetric distribution of mass within a massive disc can indeed
significantly perturb the orbit of planetesimals, causing large
eccentricities and unphased orbits, which may possibly halt the
accretion process.

\begin{figure}[hpt]
  \hskip -1 truecm
  \includegraphics[width=0.63\hsize,angle=-90]{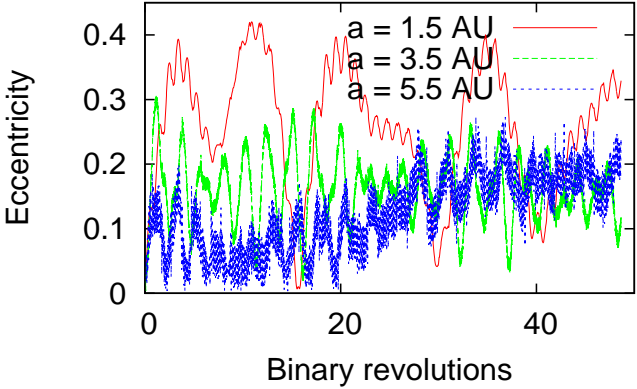}
  \caption{\label{f_pla1}Planetesimal orbital evolution without the
    gas drag force.  Large values of eccentricity are excited by the
    non-homogeneous distribution of mass within the disc. The highest
    eccentricity value is observed closer to the star where the
    companion gravitational perturbations are weaker.}
\end{figure}

To test this hypothesis we integrated the trajectories of
planetesimals orbiting within 6 AU from the primary star, and we
computed their orbital evolution in 2D under the action of (i) stellar
gravity, (ii) gas drag, and (iii) gas disc gravity. This procedure has
been implemented in previous papers, but with different assumptions
like isothermal disc models, no self-gravity and different binary
parameters \citep{paard08,kn08}.  We first present in Fig.~\ref{f_pla1}
the results of a test run without gas drag, for which we see that the
planetesimals eccentricity grows to large values regardless of their
initial location in the disc. In addition, the drift towards the inner
region of the disc, which is typical of planetesimals around single
stars, may be halted and even reversed. This drag-free case could
probably describe the evolution of large planetesimals, 100 km or
bigger, which are not significantly affected by gas drag. For the disc
and star parameters, we consider our standard case.

\begin{figure*}[hpt]
  \centering\resizebox{\hsize}{!}  {
    \includegraphics[angle=-90]{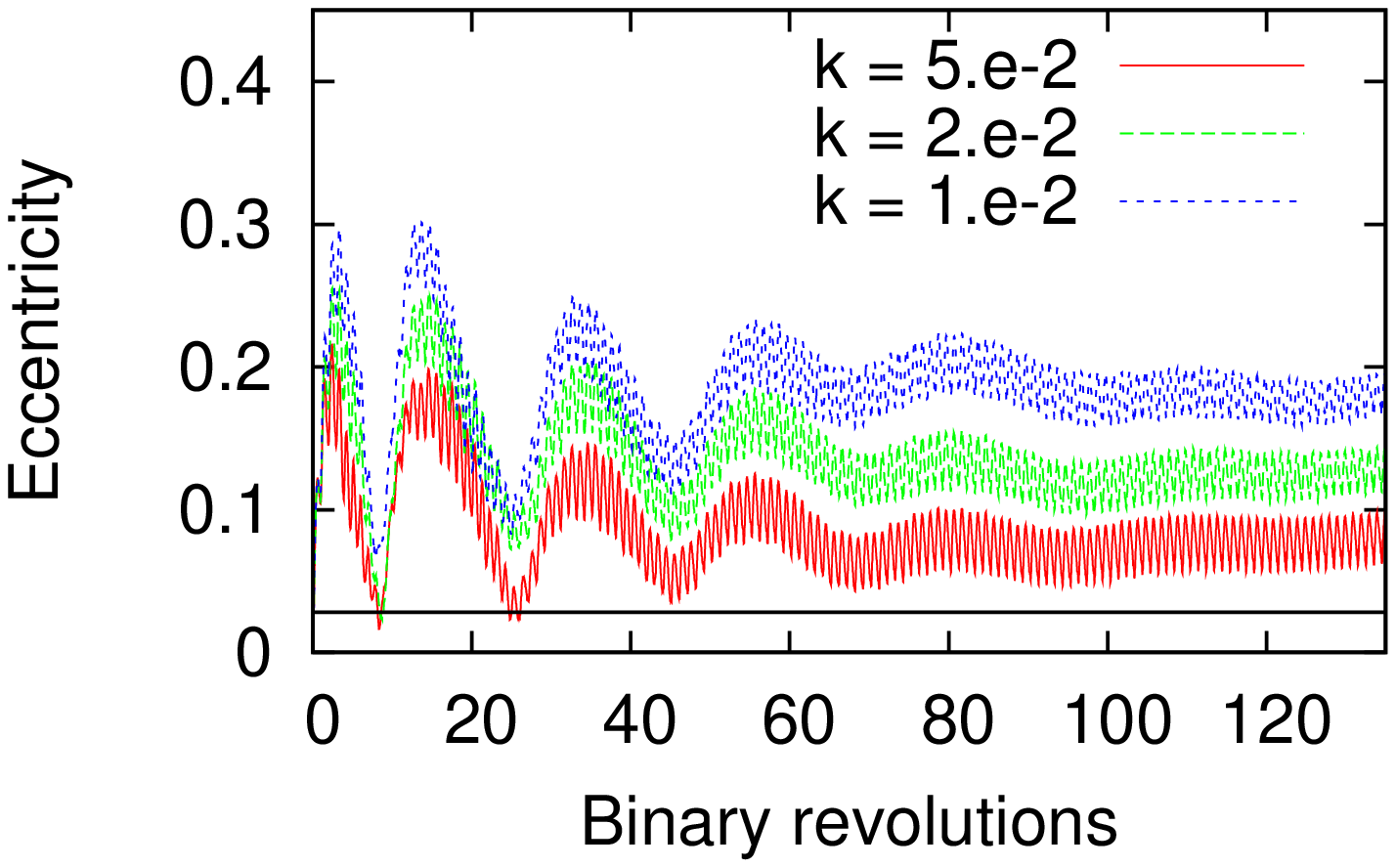}
    \includegraphics[angle=-90]{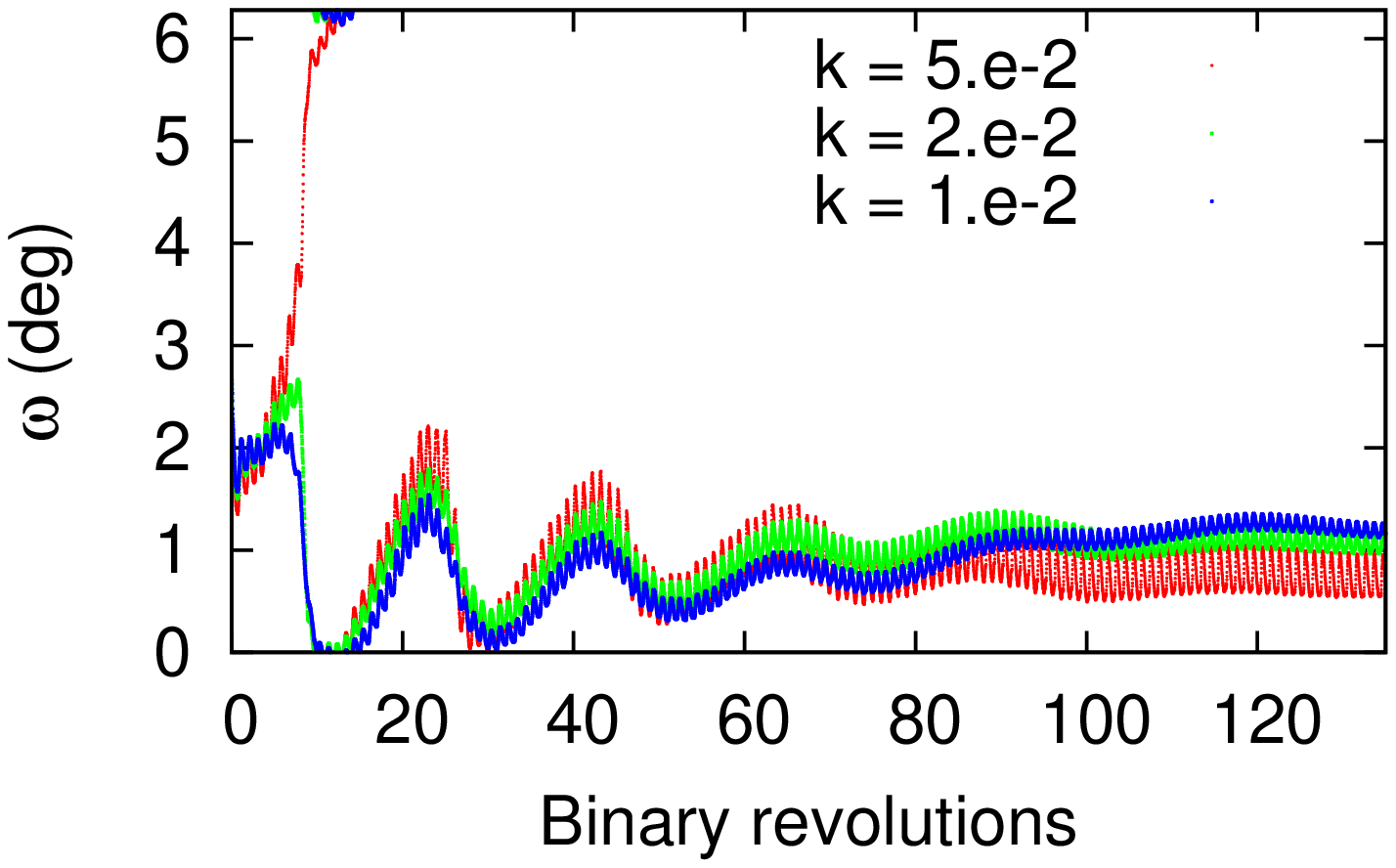}
   } 
   \centering\resizebox{\hsize}{!}  {
     \includegraphics[angle=-90]{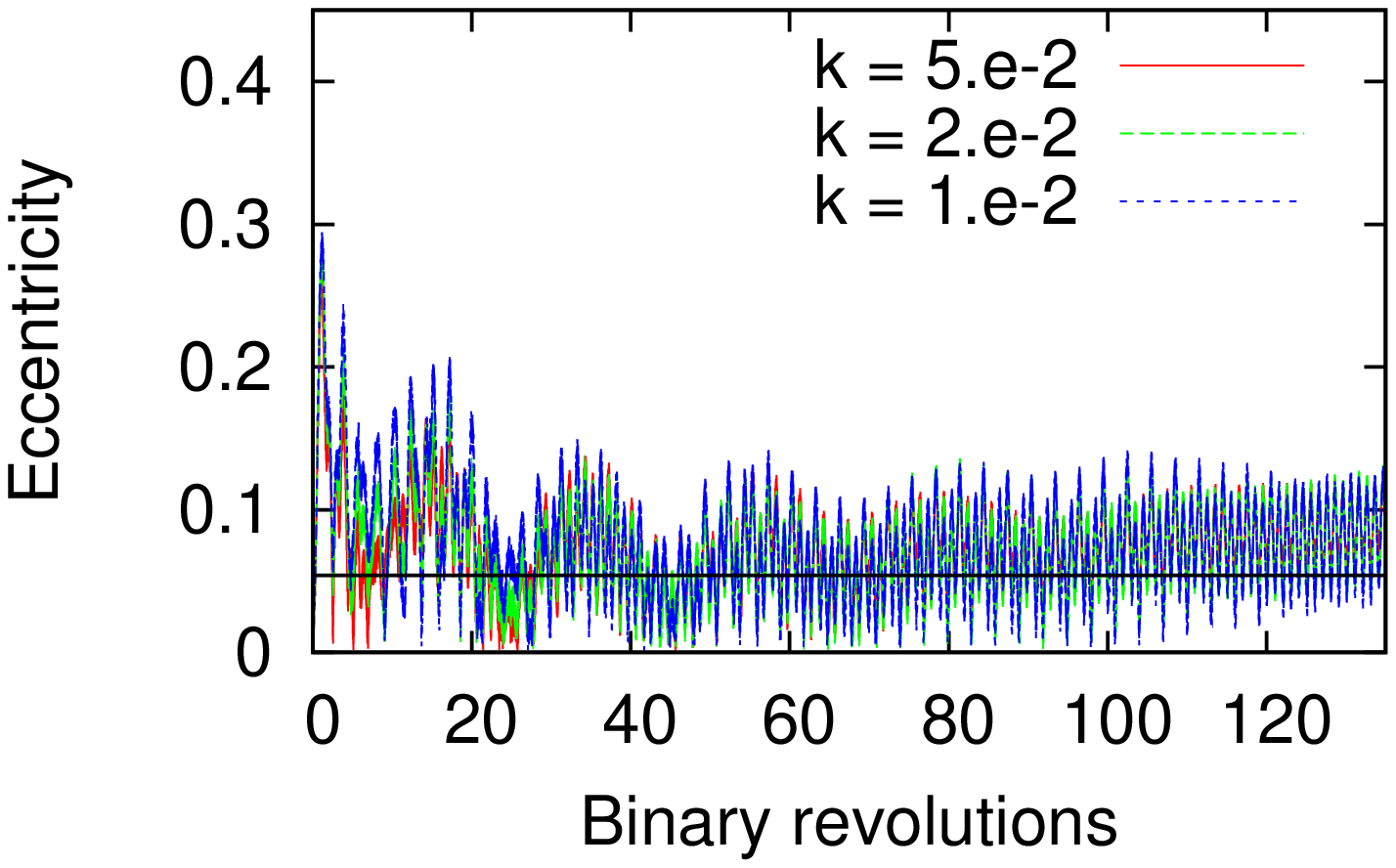}
     \includegraphics[angle=-90]{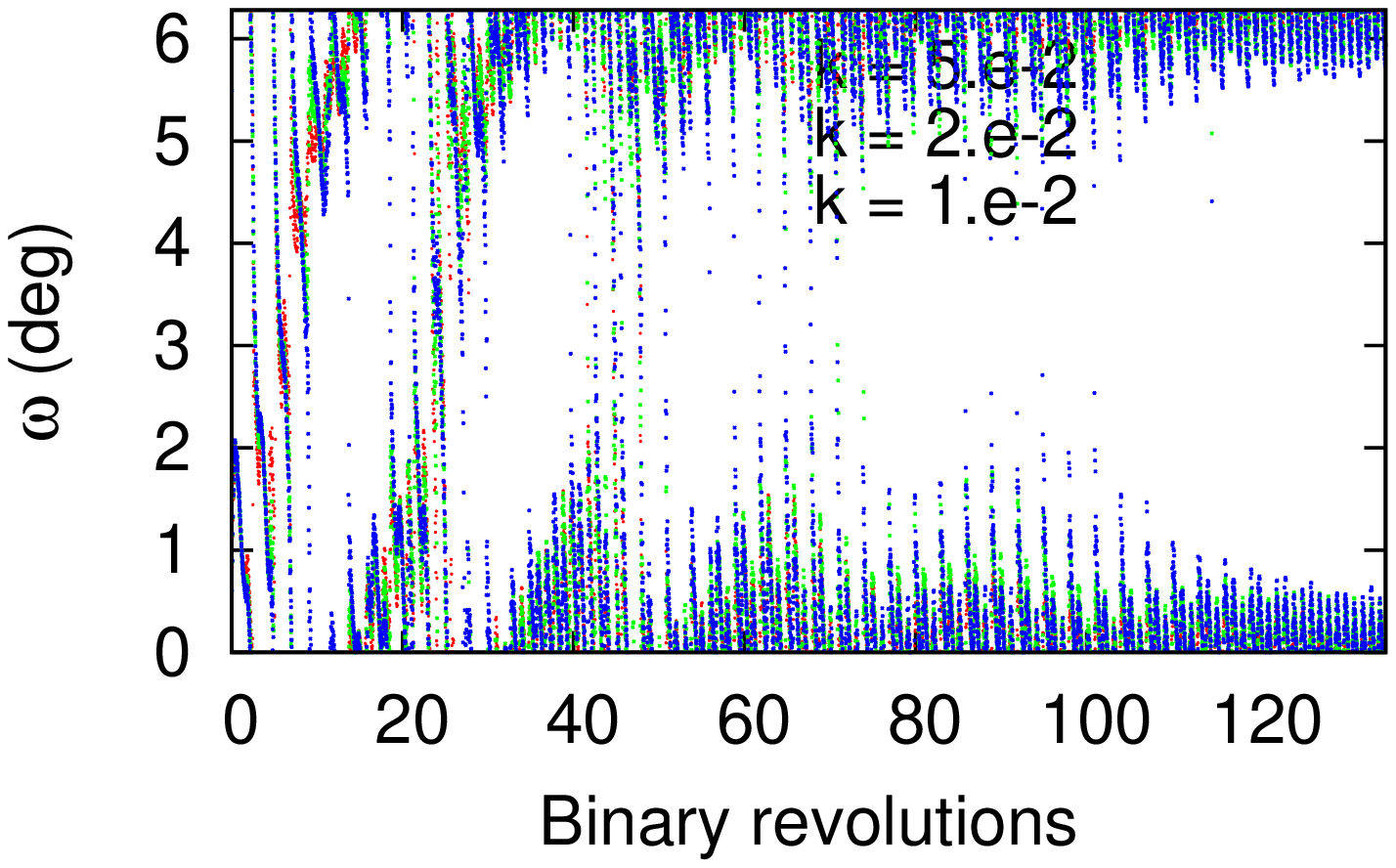}
   }
   \caption{\label{f_pla2}Orbital evolution of planetesimals with
     different sizes close to the primary star (upper plots, a = 1.5
     AU), and farther out (lower plots, a=3.5 AU). The black line on
     the left plots show the value of the forced eccentricity of the
     companion star.}
\end{figure*}

We then present the results of a full simulation including gas drag,
whose expression is given by the usual formulae ${\bf F}_{\rm d} =
K\,|{\bf v_{\rm rel}}|\,{\bf v_{\rm rel}}$, where ${\bf v_{\rm rel}} $
is the relative velocity vector of the planetesimal with respect to
the gas, and the drag parameter $K$ is equal to $\rho_{\rm g} C_{\rm
  d} / (8 \rho_{\rm pl} s)$ \citep{karo}, where $s$ is the radius of a
given planetesimal, $\rho_{\rm pl}$ its mass density, $\rho_{\rm g}$
the gas density of the protoplanetary disc, and $C_{\rm d}$ a
dimensionless drag coefficient related to the planetesimals shape
($\sim 0.4$ for spherical bodies). We explore 3 values for $K$,
corresponding to 10, 20 and 50 km-sized planetesimals in a MMSN disc,
and we consider 2 initial locations for the planetesimals: 1.5 AU and
3.5 AU from the primary.  For the 1.5 AU case, the initial strong
perturbations of the eccentric disc are partly damped by gas drag but
the steady state eccentricities are much larger than the forced
eccentricity of the companion star (see Fig.~\ref{f_pla2}, upper box).
Moreover, these steady state eccentricities do strongly vary with $K$,
i.e. with planetesimal sizes: they are almost twice as large for 50
km-sized objects as for 10 km-sized ones. As for the pericenter
longitudes, they rapidly converge to steady state values that only
marginally vary with planetesimal size.  The evolution of the 3.5 AU
case is different. The eccentricities decrease towards an equilibrium
value predicted by the balancing between the forced component of the
companion star, gas drag damping and disc gravity (see
Fig.~\ref{f_pla2}, lower plots) The steady state eccentricities are
thus lower than in the 1.5 AU case, despite being in a region where
secondary perturbations are stronger.  This result is in sharp
contrast with what was obtained by \citet{paard08} for the gas
drag-only case, where eccentricities steadily increase with increasing
semi-major axis (see Fig.~10b of that paper).  It clearly illustrates
the fact that the gas disc gravity, which is stronger in the inner and
denser part of the system, is the dominant mechanism controlling the
planetesimals' dynamical evolution. The residual short-term variations
of the eccentricity, due to the companion's perturbations, are in
contrast much larger in the 3.5 AU case than in the 1.5 AU one.  The
pericenter alignment is maintained at this distance even if it is less
collimated, as expected since the gas density is lower.

Assessing the consequences of these dynamical behaviors on the
accretion process would require estimating the distribution of impact
velocities, $v_{\rm coll}$, among the planetesimals population. It is
here not possible to directly derive $v_{coll}$ from the values of the
eccentricity because of mutual orbital phasing. The simple $v_{\rm
  coll} \propto \langle e \rangle$ relation is in this case no longer
valid and should be replaced by $v_{\rm coll} \propto \alpha_P\,
\langle e\rangle$, where the factor $\alpha_P<1$ accounts for the
phasing.  Unfortunately, proper velocity estimates are difficult to
derive with the limited number of planetesimals (50) considered
here. Simulations with at least a few thousands test particles would
be necessary \citep[e.g.][]{theb06,paard08,xie09}, but such numbers
are beyond the current computing capacities for a full model including
the gas disc's gravity. However, the preliminary results displayed in
Figs.~\ref{f_pla1} and \ref{f_pla2} seem to indicate that encounter
velocities are probably further increased by the action of disc
gravity, and this for 2 reasons: (i) Even if a significant orbital
phasing is observed (right hand plots of Fig.~\ref{f_pla2}) it is
nevertheless not perfect, especially in the 3.5 AU case. Even in the
1.5 AU case it is size-dependent, which increases the factor
$\alpha_P$ in any realistic planetesimal population with a size
distribution. (ii) The values of the steady-state eccentricities are
higher than in the gas-drag only case, especially closer to the star
where velocities are relatively low in the gas-drag only runs (compare
for example the upper-left graph of Fig.~\ref{f_pla2} to Fig.~10 of
\citet{paard08}). These steady-state eccentricities are in addition
strongly size-dependent.  These two effects separately increase each
of the two components, $\alpha_P$ and $\langle e \rangle$, controlling
the value of the encounter velocities.  The extent of this velocity
increase, and thus its concrete effect on the accretion process,
cannot be quantitatively estimated here. It will be the purpose of a
forthcoming study specifically addressing this issue. We are however
confident that the general tendency is for disc gravity to act
$against$ planetesimal accretion.

%FFFFFFFFFFFFFFFFF
%FFFFFFFFFFFFFFFFF

%FFFFFFFFFFFFFFFFF
%FFFFFFFFFFFFFFFFF

% ====================== %
\section{Concluding remarks}
% ====================== %
We have shown that the evolution of a circumstellar disc in close binary-star 
systems strongly depends on the disc's mass and thermodynamical properties. 
In locally isothermal disc models, where the temperature profile 
remains constant with time, the averaged eccentricity in the disc inner parts 
changes dramatically with varying the disc temperatures (or, equivalently, the 
disc aspect ratios $h=H/r$), as illustrated in Fig.~\ref{f_iso}. In our model, 
the disc eccentricity typically peaks at $\approx 0.4$ for $h \approx 0.05$. 
Such large eccentricities result in the formation of an internal elliptic low-density 
region in the disc. Radiative discs, on the other hand, have a  smoother density 
profile, and their eccentricity takes smaller values than in 
locally isothermal models with same temperature profile. Radiative damping of 
the waves induced by the secondary companion contributes to keep 
the disc eccentricity to a fairly small value, which in our model typically amounts to 
$\sim 0.05$ in the disc inner parts.

In both locally isothermal and radiative discs, self-gravity causes the 
libration of the disc orientation at an angle $\pi$ with respect to the
apsidal line of the binary orbit.  However, the libration is not
coherent within the disc, and the orientation of the gas streamlines changes 
with radial distance. 

The averaged disc eccentricity in radiative discs is almost insensitive 
to the binary's eccentricity, although 
it should slowly increase with time as the disc mass decreases due 
to its viscous evolution. A disc that is 10 times less massive than our 
nominal disc (that is, having $\Sigma_0 = 0.1 \Sigma_{\rm MMSN}$) has a 
large eccentricity (about 0.3 for $e_b=0.4$). This behavior is not observed in locally 
isothermal discs. 

Based on these different behaviors, we may envision an evolutionary
track for discs in binaries. In the earlier stages, the disc is massive
and probably optically thick so it is well described by radiative
models. Its eccentricity should therefore remain small, around 
0.05, and its radial density profile be smooth. Later, because of the 
viscous evolution, the disc progressively loses mass. This occurs at a fast rate
for binaries with large values of $e_b$ (see
Fig.~\ref{f_e_a}). If the disc is still optically thick, and the
radiative model is appropriate, its eccentricity is expected to grow
because of the dependence of $e_d$ with the disc mass (see
Fig.~\ref{f_mass}). If, finally, the disc becomes optically thin, 
a locally isothermal equation of state  may then be more appropriate. In this
case, the disc eccentricity will change and depend on $e_b$ as shown in
\cite{mabash}. At the same time, hotter discs will develop an internal
elliptic hole, a significant decrease in the gas density at the inner
edge of the disc.

The non-symmetric distribution of mass caused by the differential
libration of the disc orientation and the disc overall eccentric shape
causes significant perturbations to planetesimals. Their eccentricity
is driven to values significantly larger than those caused only by the
secular perturbations of the binary companion. Even when gas drag is
included, the perturbations by the eccentric disc entail large
planetesimals eccentricities, in particular in the inner zones of the
disc. It is necessary to evaluate the impact of this eccentricity on
the accumulation process of planetesimals.

% ===============
\section*{ACKNOWLEDGMENTS}
% ===============
We thank an anonymous referee for his useful comments and suggestions
that helped to improve the paper.
CB acknowledges support from a Herchel Smith Postdoctoral Fellowship.
Part of the numerical simulations were performed on the Pleiades
Cluster at U.C. Santa Cruz. We also gratefully acknowledge the
computing time provided by the Mesocentre SIGAMM machine, hosted by
the Observatoire de la C{\^o}te d'Azur.

\bibliographystyle{aa}
\bibliography{paper}

 \end{document}